\documentclass[reprint,floatfix,10pt]{revtex4-2} 
\usepackage{graphicx,amsfonts,bm,epsfig,color,latexsym,url}
\usepackage{lineno}
\usepackage{amsmath}
\usepackage{amsfonts}
\usepackage{amssymb}
\usepackage{epstopdf}
\usepackage{geometry}
\usepackage{float}
\modulolinenumbers[5]
\usepackage[bookmarks=true,bookmarksopen=true,bookmarksnumbered=true,bookmarksopenlevel=3]{hyperref}
\begin{document}
\newcommand{\bd}{\begin{document}}
\newcommand{\ed}{\end{document}}
\newcommand{\bc}{\begin{center}}
\newcommand{\ec}{\end{center}}
\newcommand{\bfr}{\begin{flushright}}
\newcommand{\efr}{\end{flushright}}
\newcommand{\lt}{\left}
\newcommand{\rt}{\right}
\newcommand{\vs}{\vspace}
\newcommand{\hs}{\hspace}
\newcommand{\ii}{\textrm{i}}

\newcommand{\beq}{\begin{equation}}
\newcommand{\eeq}{\end{equation}}
\newcommand{\bea}{\begin{eqnarray}}
\newcommand{\eea}{\end{eqnarray}}
\newcommand{\bes}{\begin{subequations}}
\newcommand{\ees}{\end{subequations}}

\newtheorem{thrm}{Theorem}[section]
\newtheorem{note}{Note}[section]
\newtheorem{dfn}{Definition}[section]
\newtheorem{ex}{Example}[section]
\newtheorem{subex}{Example}[subsection]
\newtheorem{cl}{Corrolary}[section]
\newtheorem{propo}{Proposition}[section]

\newcommand{\lb}{\linebreak}
\newcommand{\pb}{\pagebreak}
\newcommand{\mb}{\makebox}
\newcommand{\fb}{\framebox}
\newcommand{\mc}{\multicolumn}
\newcommand{\ben}{\begin{enumerate}}
\newcommand{\een}{\end{enumerate}}
\newcommand{\bit}{\begin{itemize}}
\newcommand{\eit}{\end{itemize}}
\newcommand{\un}{\underline}
\newcommand{\lefq}{\lefteqn}
\newcommand{\ba}{\begin{array}}
\newcommand{\ea}{\end{array}}
\newcommand{\beqa}{\begin{eqnarray}}
\newcommand{\eeqa}{\end{eqnarray}}
\newcommand{\beqas}{\begin{eqnarray*}}
\newcommand{\eeqas}{\end{eqnarray*}}
\newcommand{\bfg}{\begin{figure}}
\newcommand{\efg}{\end{figure}}
\newcommand{\bds}{\begin{displaymath}}
\newcommand{\eds}{\end{displaymath}}
\newcommand{\btb}{\begin{tabbing}}
\newcommand{\etb}{\end{tabbing}}
\newcommand{\para}{\parallel}
\newcommand{\pad}{\partial}
\newcommand{\nn}{\nonumber}
\newcommand{\la}{\leftarrow}
\newcommand{\ra}{\rightarrow}
\newcommand{\lgla}{\longleftarrow}
\newcommand{\lgra}{\longrightarrow}
\newcommand{\La}{\Leftarrow}
\newcommand{\Ra}{\Rightarrow}
\newcommand{\Lra}{\Leftrightarrow}
\newcommand{\Lgla}{\Longleftarrow}
\newcommand{\Lgra}{\Longrightarrow}
\newcommand{\lan}{\langle}
\newcommand{\ran}{\rangle}
\renewcommand{\a}{\alpha}
\renewcommand{\b}{\beta}
\newcommand{\g}{\gamma}
\newcommand{\G}{\Gamma}
\renewcommand{\d}{\delta}
\newcommand{\eps}{\epsilon}
\newcommand{\Th}{\Theta}
\newcommand{\s}{\sigma}
\newcommand{\lam}{\lambda}
\newcommand{\D}{\Delta}
\newcommand{\vare}{\varepsilon}
\newcommand{\pr}{\prime}
\newcommand{\ro}{\rho}
\newcommand{\nab}{\nabla}
\newcommand{\m}{\mu}
\newcommand{\n}{\nu}
\newcommand{\Sg}{\Sigma}
\newcommand{\p}{\pi}
\newcommand{\R}{I\!\!R}
\newcommand{\om}{\omega}
\newcommand{\Om}{\Omega}
\newcommand{\ze}{\zeta}
\newcommand{\vart}{\vartheta}
\newcommand{\tri}{\triangle}
\newcommand{\f}{\frac}
\newcommand{\ds}{\displaystyle}
\newcommand{\iny}{\infty}
\newcommand{\pro}{\propto}
\newcommand{\np}{\newpage}
	\title{Connected and disconnected stable regions of solitons of nonlinear Schr\"odinger equation with $\mathcal{PT}$-symmetric potential}
	\author{Niladri Ghosh}
	\altaffiliation{ Email: niladri.02mgf@gmail.com}\affiliation{Department of Mathematics, University of Kalyani, Kalyani - 741235, WB, India.}
	\author{Amiya Das}
	\altaffiliation{Email: amiya620@gmail.com}\affiliation{Department of Mathematics, University of Kalyani, Kalyani - 741235, WB, India.}
	\author{Debraj Nath}
	\altaffiliation{Corresponding author. Email: debrajn@gmail.com} \affiliation{Department of Mathematics, Vivekananda College, Kolkata - 700063, WB, India.}
\begin{abstract}
We have considered cubic nonlinear Schr\"odinger equation along with supersymmetric $\mathcal{PT}$ like potential and obtained exact stationary solutions in terms of bright and brigh-dark interacting solitons. The $\mathcal{PT}$ broken and $\mathcal{PT}$ unbroken regions are demonstrated also depicted. Connected and disconnected stable regions of bright and dark solitons are examined incorporating linear stability analysis validated by direct numerical simulations. Moreover, the strength of stability has been illustrated through excitations of bright and dark solitons.
\end{abstract}
\maketitle
\section{Introduction}
A fundamental axiom of quantum mechanics is that every physical observable is associated with a real spectrum and thus operators representing such physical observables must be Hermitian. In quantum mechanics, in order for the energy levels to be real and the theory to be probability conserving (or unitary evolution), it is usually assumed that the Hamiltonian (Schr\"odinger) operator be Hermitian. In the case of the Hamiltonian operator, this requirement gives rise to real energy levels and the theory guaranteing conservation of probability. In recent years, in a series of papers by Bender and coworkers \cite{1-19.1}, a considerable attention has been shown in a weaker version of the Hermiticity axiom in which many non-Hermitian Hamiltonians may also lead to entirely real spectra provided they possess something known as $\mathcal{PT}$ (parity-time) symmetry. The fundamental idea of $\mathcal{PT}$ symmetric quantum mechanics is to replace the concept of Hermitian Hamiltonian by weaker condition that it possess space-time reflection symmetry ($\mathcal{PT}$-symmetry). The linear space-reflection operator $\mathcal{P}$, responsible for spatial reflection, is defined through the operations $p \to -p\, , x \to -x$, while the anti-linear time-reversal operator $\mathcal{T}$ leads to $p \to -p\, , x \to x$ and to complex conjugation $\ii \to -\ii$. Dynamical systems are known to be $\mathcal{PT}$ symmetric if they remain invariant under the combined parity $\mathcal{P}$ and time-reversal $\mathcal{T}$ transformation. A necessary condition for a Hamiltonian to be $\mathcal{PT}$ symmetric is that the potential function $V_{PT} = V(x) + \ii W(x)$ should satisfy the condition $V_{PT}(x)=V^{*}_{PT} (-x)$, with $*$ denoting the complex conjugation. A $\mathcal{PT}$ symmetric system exhibits entirely real spectra as long as the robustness of the imaginary component of the complex potential is less than a certain threshold value. The corresponding system undergoes a drastic phase transition also known as spontaneous $\mathcal{PT}$ symmetry breaking \cite{makris} on exceeding the threshold value. The $\mathcal{PT}$ symmetric wave propagation in optical structures can be replaced by the complex refractive index and balanced gain-loss profile which was judiciously predicted through the experimental execution in optical metamaterials, synthetic photonic lattices, microring lasers and optically induced atomic lattices etc \cite{exp}. Though the theory was initially originated in quantum mechanics, but later on the concept of $\mathcal{PT}$ symmetry has also found applications in different fields, e.g., optics \cite{1-19.2,1-19.3,1-19.4,1-19.5,1-19.6}, electronics \cite{1-19.7}, Bose-Einstein condensation \cite{1-19.8,1-19.9,1-19.10}, metamaterials \cite{1-19.11,1-19.12,1-19.13,1-19.14,1-19.15}, etc. Furthermore $\mathcal{PT}$ symmetry has been realized experimentally \cite{1-19.16,1-19.17,1-19.18,1-19.19}.

In the last decades, the theory of existence of nonlinear localized modes in $\mathcal{PT}$ symmetric potential and their linear stability has been studied very promptly \cite{PT.NLSE,PT.NLSE2}. Several interesting potentials for example Scarf-II potential \cite{scarf}, harmonic potential \cite{harmonic}, Rosen-Morse potential \cite{Rosen-Morse}, Gaussian potential \cite{Complex.GP.pra84.2011,VariationA,CPJisha.pra2014.043855,BMidya,Mihalache.RJP61.577,Mihalache.arXiv}, sextic anharmonic double-well potential \cite{sextic},  time-dependent harmonic-Gaussian potential \cite{h-gaussian} etc are considered for study.

The nonlinear Schr\"odinger equation (NLSE) is one of the pioneer mathematical model which arises in diversified physical systems. This equation is a key model describing wave processes in plasma physics, Bose-Einstein condensates (BEC), gravitational models for quantum mechanics, wave propagation in biological and geological systems \cite{nlse1}  and nonlinear optics \cite{nlse2}. In some of these fields and many others, the NLS equation appears as an asymptotic limit for a slowly varying dispersive wave envelope propagating in a nonlinear medium. It also plays an important role in describing full spatiotemporal optical solitons or light bullets in the theory of nonlinear optics \cite{nlse2}.
During the past few years, exact solutions of NLSE with $\mathcal{PT}$ symmetric potentials and their stability have been studied by many authors since they are useful in different contexts \cite{Rosen-Morse,23-42.23,23-42.24,23-42.25,23-42.26,23-42.27,23-42.28,23-42.29,23-42.30,23-42.31,23-42.33,23-42.34,23-42.35,23-42.36,23-42.37,23-42.38,23-42.39,nath.csf,pla,nath16,nath17}. In Ref.\cite{pla}, the exact stationary solutions of derivative nonlinear Schr\"odinger equation have been obtained in presence of a $\mathcal{PT}$ symmetric potential as a sum of super-Gaussian and parabolic potentials. The theory of supersymmetry and deformed supersymmetry are used in \cite{nath16,nath17} to retrieve exact analytical localized solutions in context of a number of complex $\mathcal{PT}$ symmetric potentials and supersymmetric potentials along with power law nonlinearity.

As we know that, a pair of potentials $V_{\pm}(x)$ are said to be supersymmetric, if there exists a function $U(x)$ such that, $V_{\pm}(x)=U^2(x)\pm U'(x)$ \cite{cooper}. In this paper, we have considered $(a,v_0,w_0,U(x))$, where $U(x)=U_R(x)+\ii U_I$ such that,
\beq\label{vw}
\ba{l}
V=v_0\,Re(U^2-U^\prime)=v_0\left(U_R^2-U_I^2-U_R^\prime\right),\\
W=w_0\,Im\left(U^2-U^\prime\right)=w_0\,\left(2U_RU_I-U_I^\prime\right),
\ea
\eeq
$v_0$ and $w_0$ are amplitudes of real and imaginary parts of the potential, with $w_0\ne 0$ and $a$ is a real constant, which will be conneted to solutions of a NLSE. Then the functions $U_R$ and $U_I$ satisfy $\mathcal{PT}$-symmetric conditions
\beq\label{ptcond1}
U_R(x)=-U_R(-x),~~~~U_I(x)=U_I(-x),
\eeq
and a cubic nonlinear Schr\"odinger equation has exact solutions for deformed supersymmetric potential $V(x)+iW(x)$. The optical solitons are classified into two forms based on the sign of group velocity dispersion. Solitons in negative dispersion regime are known as bright solitons (BS) and solitons in positive dispersion regime are called dark solitons (DS). Dark soliton can be stressed as a localized pulse appearing as an acute dip on the background of continuous wave. On the other hand, bright soliton appears as a crest above the continuous wave. Recently, in a series of paper \cite{DS}, the existence of dark solitons are verified experimentally in the area of nonlinear fiber optics, plasma, BEC and waveguide arrays. In most of the cases, cubic NLSE has either a bright or dark soliton. In this paper, we will find two solutions of a cubic NLSE and they will be the ground and first excited states of linear Schr\"odinger equiation. In this paper, we will investigate stability analysis of BS and DS of a cubic nonlinear Schr\"odinger equation with deformed supersymmetric potential. Then we will find stable regions of BS and DS and regions will be verified by different numerical methods. Finally, regions will be checked by excitations of BS and DS of a cubic time dependent NLSE. 

The rest of the paper is structured as follows. In Sec. \ref{sec2.Method}, we will obtain exact stationary solutions in terms of BS and bright-dark soliton interaction of a cubic nonlinear Schr\"odinger equation along with complex $\mathcal{PT}$ symmetric potential. In Sec. \ref{sec3.linear}, linear stability analysis is illustrated incorporating direct numerical simulations, which assures about stable behavior of BS and DS depending on potential parameters. Also, excitations of nonlinear BS and DS of a time dependent NLSE and some numerical results have been reported in Sec. \ref{sec4.excitation}. Finally, conclusions have been drawn in Sec. \ref{sec5.con}.
\section{Methodology}\label{sec2.Method}
Let us consider a cubic NLSE
\beq\label{nlst}
i\Psi_t=-\Psi_{xx}+\left(V(x)+\ii W(x)\right)\Psi+g|\Psi|^2\Psi
\eeq
with a complex $\mathcal{PT}$-symmetric potential $V(x)+\ii W(x)$, where $g$ is the nonlinearity parameter. To solve Eq.(\ref{nlst}) a transformation
\beq\label{mu}
\Psi(x,t)=e^{-\ii \mu t}\psi(x)
\eeq
is considered, where $\mu$ is a real propagation constant. Then from Eq.(\ref{nlst}) one can obtain
\beq\label{nls}
-\f{d^2\psi}{dx^2}+(V+\ii W)\psi+g|\psi|^2\psi=\mu\psi
\eeq
and it has a solution of the form $C_0\,e^{-\int (U_R+\ii \,a\,U_I)dx}$, $a\in \mathbb{R}$. In particular if, $(a,v_0,w_0,U)=(1,1,1,U)$ and $g=0$, then the NLSE reduces to a LSE with exact supersymmetric potentials with zero energy normalized solution $e^{-\int (U_R+\ii \,U_I)dx}$. In this paper we will find solutions of NLSE (\ref{nls}) as well as (\ref{nlst}), where $(a,v_0,w_0)\ne (1,1,1)$.
\subsection{Solution 1}
A solution of Eq.(\ref{nls}) is taken to be of the form
\beq\label{sol}
\psi_0=C_0\,e^{-\int (U_R+\ii \,a\,U_I)dx},~~\mu=\mu_0
\eeq
where $C_0$ is the amplitude and $\lim\limits_{|x|\rightarrow\infty}|\psi_0(x)|=0$. Now substituting Eq.(\ref{sol}) into Eq.(\ref{nls}) and equating real and imaginary parts, we obtain \cite{nath16}
\beq
\ba{l}\label{relation}
\left(v_0-1\right)\left(U_{R}^{2}-U_{R}^{'}\right)+\left(a^2-v_0\right)U_{I}^{2}\\
+gC_{0}^2\,e^{-2\int U_{R}(x)\,dx}=\mu_0,\\
\left(2U_{R}U_{I}-U_{I}^{'}\right)\left(w_0-a\right)=0.
\ea
\eeq
If $2U_RU_I-U'_I=0$, then $\mathcal{PT}$-symmetric potential becomes real potential. In this paper we have considered $w_0=a$ and therefore $(a,v_0,w_0,U)$ exists which satisfy relations of (\ref{relation}) and solution (\ref{sol}) esists.
\subsection{Solution 2}
If possible let, there be another solution of the form
\beq\label{nthsolution}
\ba{l}
\psi_1(x)=C_1\,\ds e^{-\ds\int\left(U_R+\ii \,a\,U_I\right)\,dx}F(x),\mu=\mu_1,
\ea
\eeq
where $\lim\limits_{|x|\rightarrow\infty}|\psi_1(x)|\ne0$.
Using Eq.(\ref{nthsolution}) into Eq.(\ref{nls}), we obtain
\beq\label{relationpsi1}
\ba{ll}
-F''+2\left(U_R+\ii \,a\,U_I\right)F'\\
+g\left\{C_1^2|F|^2-C_0^2\right\}\,e^{-2\ds\int U_R\,dx}F=\left(\mu_1-\mu_0\right)F.
\ea
\eeq
Eq.(\ref{relationpsi1}) is the additional condition for existence of solution of Eq.(\ref{nthsolution}). In particular, $g=0$, then conditions are
\beq
\ba{l}\label{condition.LSE}
\left(v_0-1\right)\left(U_{R}^{2}-U_{R}^{'}\right)+\left(a^2-v_0\right)U_{I}^{2}=\mu_0,\\
\left(2U_{R}U_{I}-U_{I}^{'}\right)\left(w_0-a\right)=0,\\
-F''+2\left(U_R+i\,a\,U_I\right)F'=\left(\mu_1-\mu_0\right)F.
\ea
\eeq
Eq.(\ref{condition.LSE}) has the solution $(a,v_0,w_0)=(1,1,1)$ and $F$ satisfies the relation
\beq\label{Eq.F}
-F''(x)+2U(x)F'(x)=\mu_1\,F(x),
\eeq
which has infinitely many solutions depends on $U(x)$. In this paper, we will find solution of (\ref{relationpsi1}) by an example. Then the NLSE (\ref{nlst}) has two solutions of the forms (\ref{sol}) and (\ref{nthsolution}). The solution (\ref{solpsi0}) is a BS and the solution (\ref{solpsi1}) may be BS or DS. 
\subsection{Example}
\begin{figure}[h]
	\includegraphics[width=10cm,height=8cm]{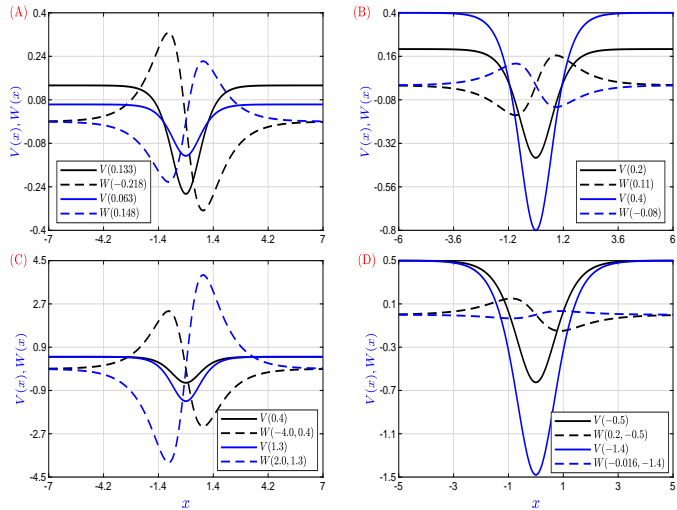}
	\caption{\label{Fig1.potential} Plot of real and imaginary parts of $\mathcal{PT}$-symmetric potential (\ref{VWEx1}). Parameters are (A) black lines $v_0=0.133,\b=1, w_0=-0.218$ and blue lines $v_0=0.063,\b=1, w_0=0.148$; (B) black lines $v_0=0.2,\b=1, w_0=0.11$ and blue lines $v_0=0.4,\b=1, w_0=-0.08$; (C) black lines $v_0=0.5,\b=0.4, w_0=-4$ and blue lines $v_0=0.5,\b=1.3, w_0=2$; (D) black lines $v_0=0.5,\b=-0.5, w_0=0.2$ and blue lines $v_0=0.5,\b=-1.4, w_0=-0.016$.}
\end{figure}

\begin{figure}[h]	
	\includegraphics[width=10cm,height=8cm]{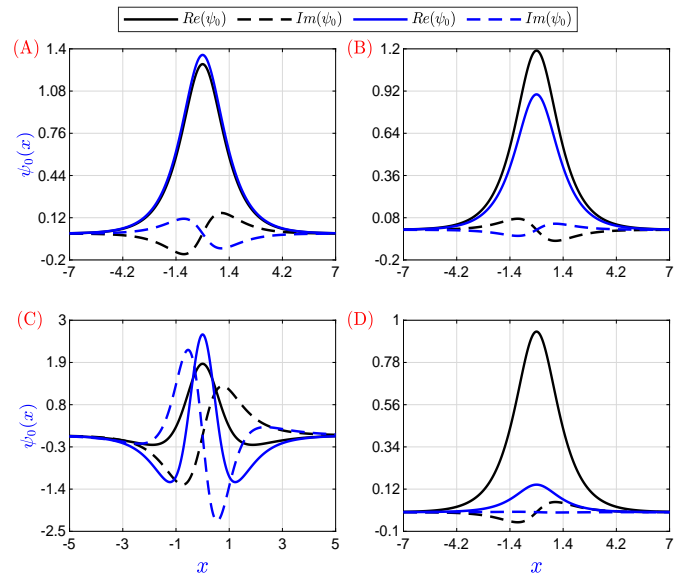}
	\caption{\label{Fig2.solution1} Plot of real and imaginary parts of (\ref{solpsi0}). The parameters are taken from (A) stable $\mathcal{PT}$ broken region: for black lines $v_0=0.133,w_0=-0.218$ and for blue lines $v_0=0.063,w_0=0.148$; (B) stable $\mathcal{PT}$ unbroken region: for black lines $v_0=0.2,w_0=0.11$ and for blue lines $v_0=0.4,w_0=-0.08$; (C) unstable $\mathcal{PT}$ broken region: for black lines $\b=0.4,w_0=-4.0$ and for blue lines $\b=1.3,w_0=2.0$, (D) stable $\mathcal{PT}$ unbroken region: for black lines $\b=-0.5,w_0=0.2$ and for blue lines $\b=-1.4,w_0=-0.016$. All regions are shown in Fig. \ref{Fig6.solution1g-1phasetransition}.}
\end{figure}

\begin{figure}[h]	
	\includegraphics[width=10cm,height=8cm]{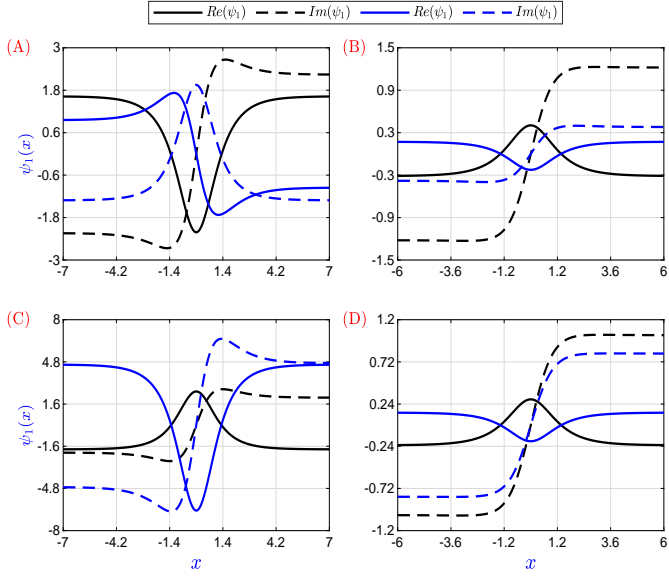}
	\caption{\label{Fig3.solution2} Plot of real and imaginary parts of (\ref{solpsi1}). The parameters are taken from (A) unstable $\mathcal{PT}$ broken region: (black lines) $v_0=-0.2,w_0=0.40$ and (blue lines) $v_0=0.4,w_0=-0.45$; (B) stable $\mathcal{PT}$ unbroken region: (black lines) $v_0=0.20,w_0=-0.16$ and (blue lines) $v_0=0.65,w_0=0.27$; (C) unstable $\mathcal{PT}$ broken region: (black lines) $\b=0.23,w_0=-2.00$ and (blue lines) $\b=0.29,w_0=1.70$, (D) stable $\mathcal{PT}$ unbroken region: (black lines) $\b=-0.10,w_0=1.40$ and (blue lines) $\b=-0.85,w_0=-0.13$ . All the regions are shown in Fig. \ref{Fig8.solution2g1phasetransition}.}
\end{figure}

Let us choose $U_R(x)=\tanh x,~~ U_I(x)=\b\rm{sech }\,x$, where $\b$ is a real constant.
Then $\mathcal{PT}$-symmetric potential $V(x)+iW(x)$ is defined by
\begin{equation}\label{VWEx1}
\ba{ll}
V(x)=v_0-v_0(2+\b^2)\,sech^2x,\\
W(x)=3w_0\b\,sech\,x\,\tanh x.
\ea 
\end{equation}
This potential is known as $\mathcal{PT}$-symmetric Scarf potential \cite{cooper}. In Fig. \ref{Fig1.potential}, we have plotted real and imaginary parts of the potential, for different values of parameters and the parameters are taken from $\mathcal{PT}$ broken and unbroken region of the linear Schr\"odinger equation with this complex potential. In Sec. \ref{linear}, the $\mathcal{PT}$ broken and unbroken symmetry are discussed. It is to be noted that, this potential was considered within the context of cubic NLSE  \cite{1-19.2,23-42.23,23-42.24,23-42.26,Rosen-Morse,23-42.33,23-42.37,23-42.39,43,nath16,nath17}.
Then solution 1 is given by
\beq\label{solpsi0}
\Psi_0(x,t)=C_0\,\rm{sech}\,x\,e^{-\ii w_0\b\tan^{-1}\left(\sinh x\right)-\ii \mu_0 t},
\eeq
and solution 2 is  given by
\beq\label{solpsi1}
\Psi_1(x,t)=C_1\left(\g \rm{sech}\,x+\ii\d\tanh x\right)e^{-\ii\,w_0\b\tan^{-1}\left(\sinh x\right)-\ii \mu_1 t},
\eeq
where
\beq\label{condition.solpsi0}
~~\mu_0=v_0-1,~~C_0=\left\{\f{v_0(2+\b^2)-w_0^2\b^2-2}{g}\right\}^{\f{1}{2}},
\eeq
and
\beq\label{condition.solpsi1}
\ba{ll}
\mu_1=v_0+\ds\f{v_0(2+\b^2)-w_0^2\b^2-2}{4\b^2w_0^2-1},\\
C_1=\ds\left\{\f{v_0(2+\b^2)-w_0^2\b^2-2}{g\d^2(4\b^2w_0^2-1)}\right\}^{\f{1}{2}},\\
\g=-2\b\d w_0.
\ea
\eeq
We have plotted real and imaginary parts of solution (\ref{solpsi0}) in Fig. \ref{Fig2.solution1} and of solution (\ref{solpsi1}) in Fig. \ref{Fig3.solution2} for different set of parameters. It is to be noted that, for Fig. \ref{Fig2.solution1} and Fig. \ref{Fig3.solution2}, all parameters are taken from $\mathcal{PT}$ broken stable, $\mathcal{PT}$ broken unstable and $\mathcal{PT}$ unbroken stable, $\mathcal{PT}$ unbroken unstable regions of Fig. \ref{Fig6.solution1g-1phasetransition}, \ref{Fig7.solution1g1phasetransition} and \ref{Fig8.solution2g1phasetransition}. We observe that, solution (\ref{solpsi0}) is a BS and solution (\ref{solpsi1}) is a BS, if $w_0^2>\f{1}{4\b^2}$ and a DS, if $w_0^2<\f{1}{4\b^2}$. Therefore, the phase transition between BS and DS is occured for the solution (\ref{solpsi1}) at the points on the curve $w_0^2=\f{1}{4\b^2}$. Now, we will focus on LSE with complex potential (\ref{VWEx1}).
\subsection{Spectrum of linear Schr\"odinger equation}\label{linear}
\begin{figure}[h]	
	\includegraphics[width=8cm,height=5cm]{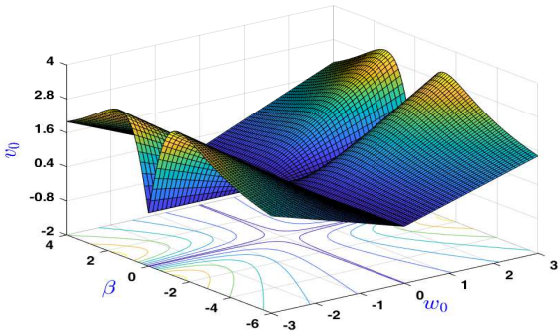}
	\caption{\label{Fig4.phase} Plot of $\mathcal{PT}$-symmetric phase diagram of $v_0$ of the linear operator (\ref{linearspectrum}) with respect to $w_0$ and $\beta$.}
\end{figure}

\begin{figure}[h]
	\includegraphics[width=10cm,height=10cm]{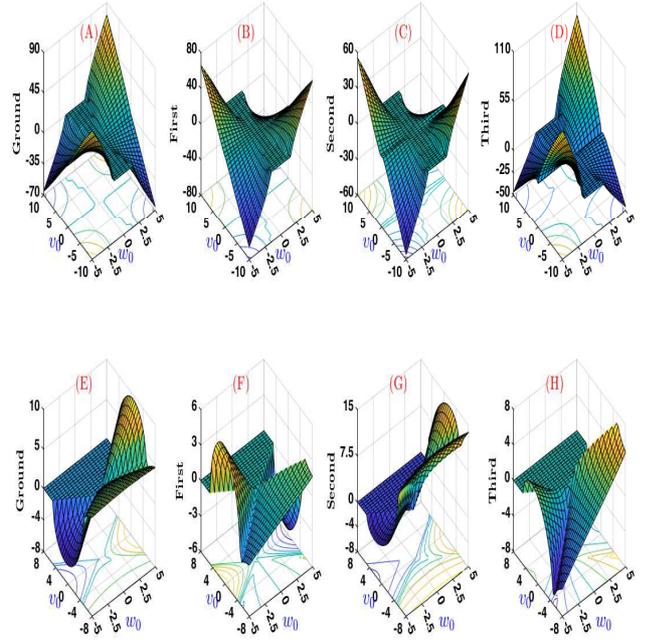}
	\caption{\label{Fig5.eigenvaluef5} Plot of imaginary parts of some eigen values of (\ref{linearspectrum}), for (A)-(D) $v_0=0.5$ and for (E)-(H) $\b=1$.}
\end{figure}

We first consider the linear spectrum of $L$ with $\mathcal{PT}$-symmetric potential (\ref{VWEx1}) as
\beq\label{linearspectrum}
\ba{l}
L\psi(x)=\mu\psi,\\
L\equiv -\f{d^2}{dx^2}+V(x)+\ii W(x),
\ea
\eeq
where $\mu$ is the eigenvalue and $\psi(x)$ is the corresponding eigenfunction. For the bound states  $\lim\limits_{x\rightarrow\infty}|\psi(x)|\rightarrow 0$. Now spectrum of the linear operator $L$ with complex $\mathcal{PT}$-symmetric potential (\ref{VWEx1}) may be real or complex conjugates. Then, one can find $v_0,w_0,\b$, for which linear operator has real spectrum and it is possible, if \cite{zafarahmed}
\beq
3|w_0\b|\le \f{1}{4}+v_0(2+\b^2).
\eeq
Fig. \ref{Fig4.phase} exhibits $\mathcal{PT}$ symmetry broken and unbroken phase transition of the complex potential (\ref{VWEx1}). The region on and above the surface of Fig. \ref{Fig4.phase} is $\mathcal{PT}$ unbroken and the region below the surface is $\mathcal{PT}$ broken. 
The imaginary parts of some eigen values of the linear operator $L$ (\ref{linearspectrum}) are shown in Fig. \ref{Fig5.eigenvaluef5}, in $(w_0,v_0)$ plane for $\b=1$ and in $(w_0,\b)$ plane for $v_0=0.5$. From Fig. \ref{Fig4.phase} one can find a suitable set of parametrs for which the potential is $\mathcal{PT}$ unbroken and the spectrum of operator $L$ \ref{linearspectrum} are real. First of all under this set of parameters, we will investigate the linear stability analysis of solutions (\ref{solpsi0}) and (\ref{solpsi1}) and we denote this point by $P_1$ and then we will invesigate further for another set of parameters, which is denoted by $P_2$. Next, we will investigate their stability analysis for collection of points joing $P_1$ and $P_2$ by a curve in $(w_0,v_0)$ plane for fixed $\b$ and in $(w_0,\b)$ plane for fixed $v_0$ in the next Sec. \ref{sec3.linear}. 
\section{Linear stability analysis}\label{sec3.linear}
One of the most important property of solutions of NLSE is their stability. Here we will examine linear stability of solutions of a NLSE. To this end, solution in Eq.(\ref{mu}) is given small perturbations $v(x), w(x)$ and is taken to be of the form \cite{jybook,nixon,zezyulin.epl,nath.optcom,nath.epjp}
\beq \label{per}
\ba{l}
\Psi(x,t)=\{\psi(x)+\left[\widetilde{v}(x)+\widetilde{w}(x)\right]e^{\lam
	t}\\
+\ds\left[\widetilde{v}^{*}(x)-\widetilde{w}^{*}(x)\right]e^{\lam^{*}t}\}e^{\ii \mu t},
\ea
\eeq
where the superscript $*$ represents the complex conjugation, $|v|,|w|\ll 1$ are infinitesimal normal mode perturbation eigenfunctions, which may grow during propagation with the perturbation growth rate $\lambda$. Now substituting the above expression in Eq.(\ref{nlst}) and linearizing, one gets the following coupled set of eigenvalue problem:
\beq\label{lameig}
\ii\left(\ba{cc}h_0&\nabla^2 +h_1\\
\nabla^2+h_2&h_3\ea\right)\left(\ba{c}\widetilde{v}\\\widetilde{w}\ea\right)=\lam
\left(\ba{c}\widetilde{v}\\\widetilde{w}\ea\right),
\eeq
where
\beq \ba{l}
h_0=\ds\f{g}{2}\left(\psi^2-\psi^{*2}\right)+i~W(x),\\
h_1=-\mu+V(x)+g|\psi|^2+\ds g\left\{|\psi|^2-\f{1}{2}\left(\psi^2+\psi^{*2}\right)\right\},\\
h_2=-\mu+V(x)+g|\psi|^2+\ds g\left\{|\psi|^2+\f{1}{2}\left(\psi^2+\psi^{*2}\right)\right\},\\
h_3=-\ds\f{g}{2}\left(\psi^2-\psi^{*2}\right)+i~W(x).
\ea
\eeq
The linear stability is determined by the nature of the eigenvalue problem (\ref{lameig}). If there exists any $\lambda$ with a positive real part, then the perturbed solution will grow exponentially with $t$ resulting in unstable mode. On the other hand, mode is completely stable only when, real parts of $\lambda$ are not positive. The BS and DS are satisfying boundary conditions
\beq\label{boundary.bs}
\ba{l}
|\Psi_0(x,t)|\rightarrow 0, |x|\rightarrow \infty,~t>0,\\
\f{\partial}{\partial x} \Psi_0(x,t)\rightarrow 0, |x|\rightarrow \infty,~t>0,
\ea 
\eeq
\beq\label{boundary.ds}
\ba{l}
|\Psi_1(x,t)|\rightarrow C_1|\d|, |x|\rightarrow \infty,~t>0, w_0^2<\f{1}{4\b^2},\\
\f{\partial}{\partial x} \Psi_1(x,t)\rightarrow 0, |x|\rightarrow \infty,~t>0.
\ea 
\eeq 
In this paper, we will apply the  Fourier collocation method (FCM) and finite difference method (FDM) for finding eigenvalues of Eq.(\ref{lameig}) for BS and DS respectively. The simulation of these two solutions are obtained by pseudospectral method and finite difference method. Using the definition of power and the VK statbility condition \cite{VK}, one can say that, BS defined in (\ref{solpsi0}) is stable, if $g>0$ and unstable, if $g<0$, for the real potential i.e., when $w_0=0$. The stable region of BS and DS can not be defined for arbitrary values of $v_0,w_0$ and $\b$. Therefore, we will find stable regions in $(w_0,v_0)$ plane for fixed $\beta$ and in $(w_0,\b)$ plane for fixed $v_0$ by applying some numerical techniques. 
\subsection{Numerical method}
All the special grid points are defined by $x_j=-L+\delta x j$, $j=1,2,3,...,N+1$ ($L$ being the half-width), where $\delta x=2L/N$ is taken to be the lattice spacing (resolution). The left and right boundary points are denoted by $j=1$ and $j=N+1$, respectively. The boundary conditions for BS and DS at the end points are given by 
\beq
\ba{ll}\label{boundary.bs.L}
\ds|\Psi_0(\pm L,t)|= 0,~\ds\left[\partial_x\Psi_0(x,t)\right]_{x=\pm L}=0,
\ea 
\eeq
and
\beq\label{boundary.ds.L}
\ba{l}
\ds|\Psi_1(\pm L,t)|=C_1|\d|,
\ds\left[\partial_x\Psi_1(x,t)\right]_{x=\pm L}=0, w_0^2<\f{1}{4\b^2}.
\ea 
\eeq 
In this paper, we will consider two finite difference methods for discretizing Eqs. (\ref{nlst}), (\ref{boundary.bs.L}) and (\ref{boundary.ds.L}). For BS (\ref{solpsi0}), we will apply (i) speudospectral method for second-order spatial derivatives and 4th order Runge-Kutta (RK4) for the temporal derivative and (ii) second-order central-difference formula for second-order spatial derivatives and Crank-Nicholson finite difference method for the temporal derivative. Similarly, for DS we will apply (iii) second-order central-difference formula for second-order spatial derivatives and 4th order Runge-Kutta (RK4) for the temporal derivative and (iv) second-order central-difference formula for second-order spatial derivatives and Crank-Nicholson finite difference method for the temporal derivative \cite{Crank-Nicolson}. Then, a solution will be considered a stable mode, if none of eigen frequencies $\lam=\lam_R+i\lam_I$ has a positive real part $\lam_R$. Finally, the stable regions will be obtained from the eigenvalue problem (\ref{lameig}) and they will be checked by direct dynamical evolution of the NLSE (\ref{nlst}) forward with respect to $t$. To validate linear stability, we will add random noise perturbation for stable mode add eigen vector $(\widetilde{v},\widetilde{w})^T$ of (\ref{lameig}), which corresponds to largest positive real part $\lam_R$ for unstable mode to initial solution. The random numbers are generated from $(0,1)$. 
\subsection{Stability of BS}
\begin{figure}[h]
	\centering
	\includegraphics[width=8cm,height=5cm]{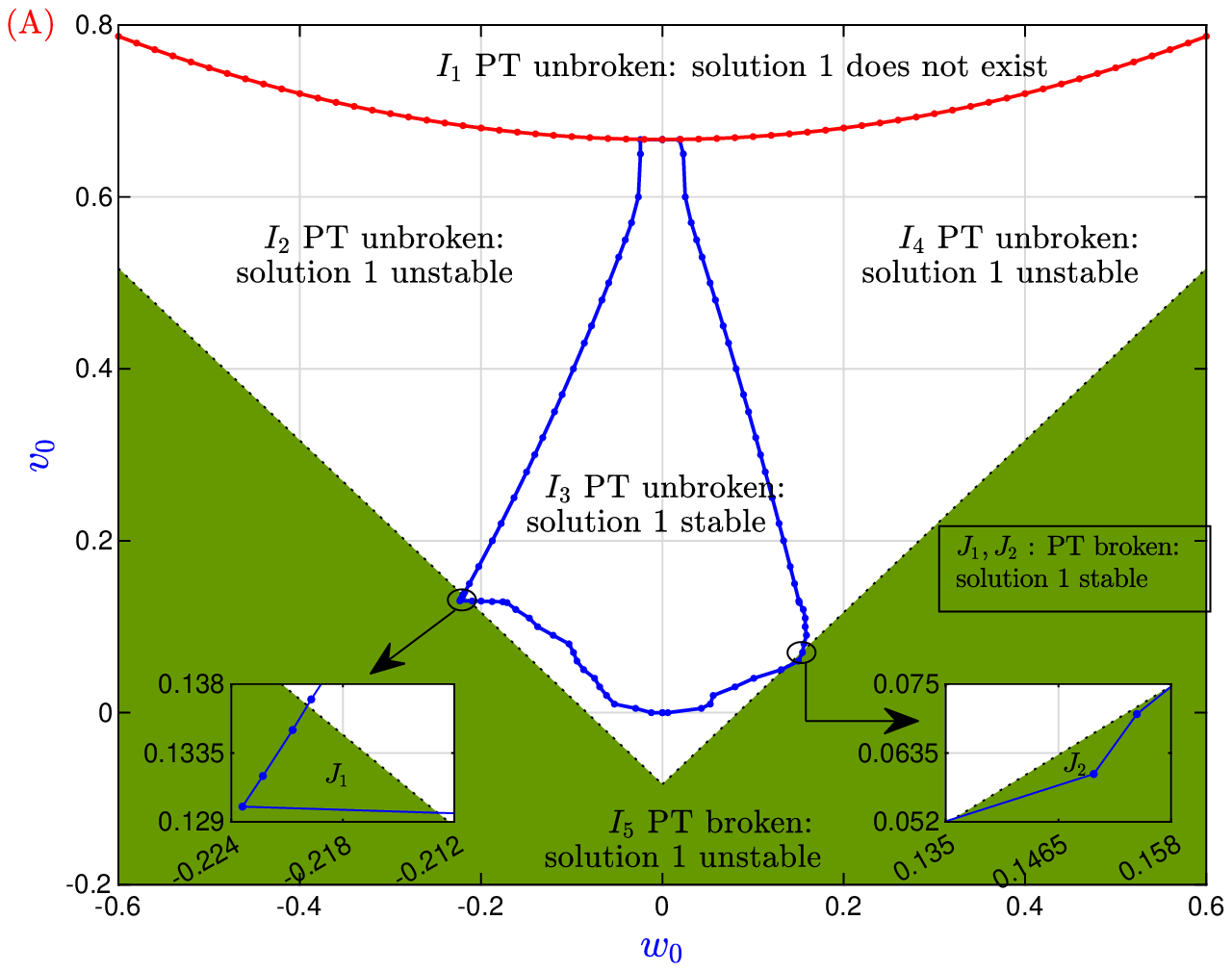}\\
	\includegraphics[width=8cm,height=5cm]{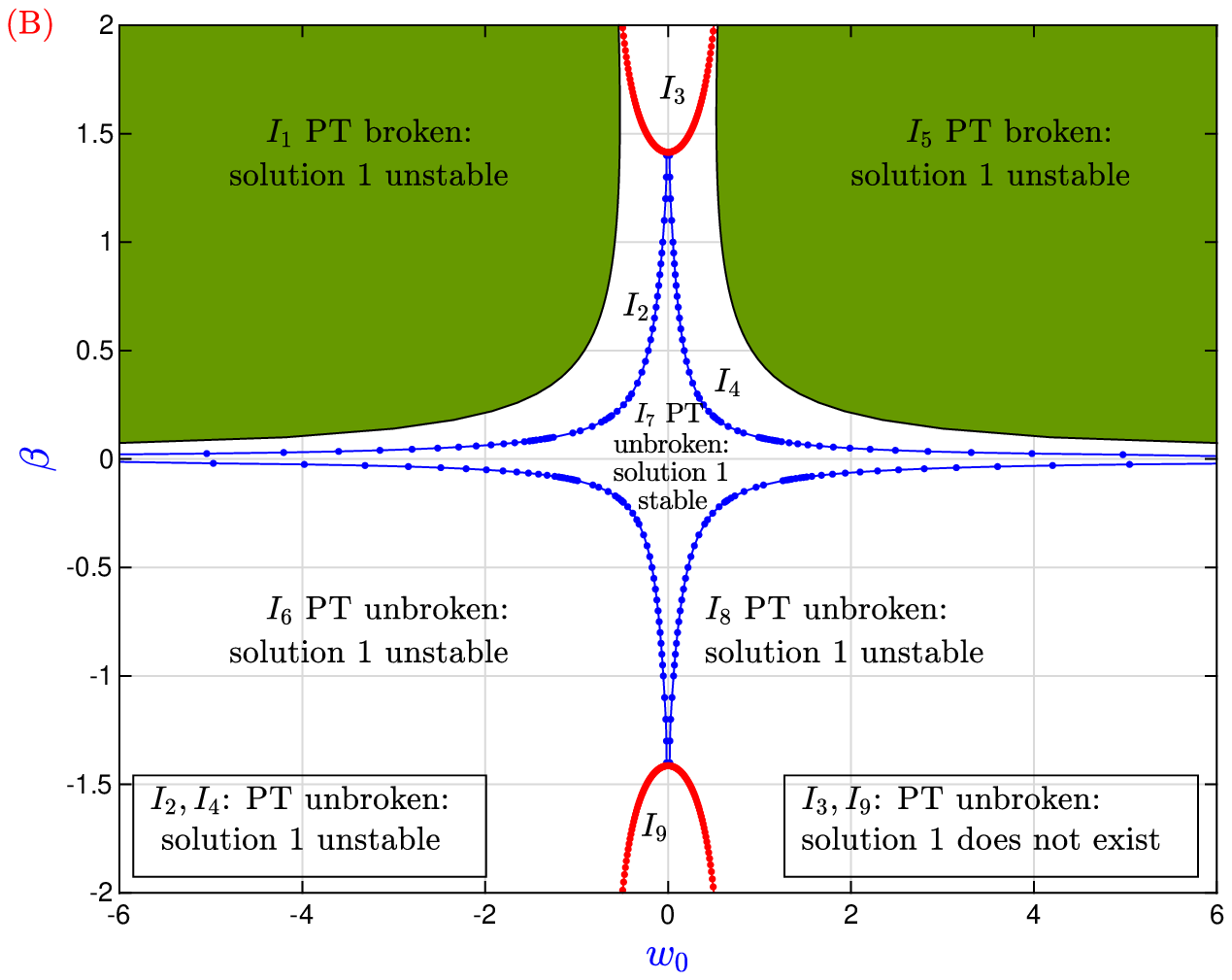}
	\caption{\label{Fig6.solution1g-1phasetransition}(A) $\mathcal{PT}$-symmetric phase transitions of the linear operator (\ref{linearspectrum}) with potential (\ref{VWEx1}) and stable-unstable regions of the solution (\ref{solpsi0}) in $(w_0,v_0)$ plane, for $g=-1,\b=1.0$. (B) The phase transitions for the same linear operator in $(w_0,\b)$ plane and stable-unstable mode of the solution (\ref{solpsi0}) in $(w_0,\b)$ plane, for $g=-1,v_0=0.5$.}
\end{figure}

\begin{figure}[h] 
	\centering
	\includegraphics[width=10cm,height=5cm]{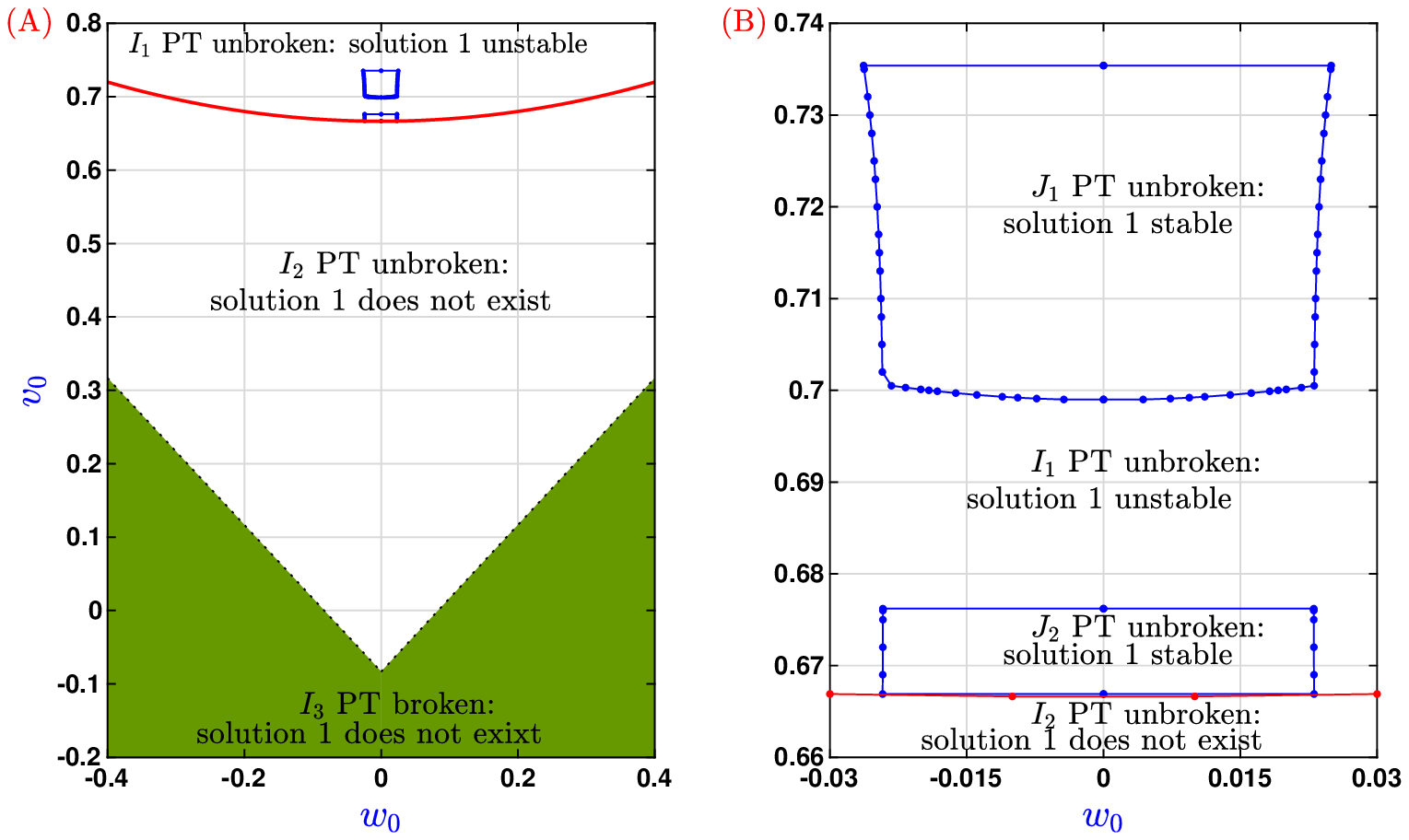}\\
	\includegraphics[width=10cm,height=5cm]{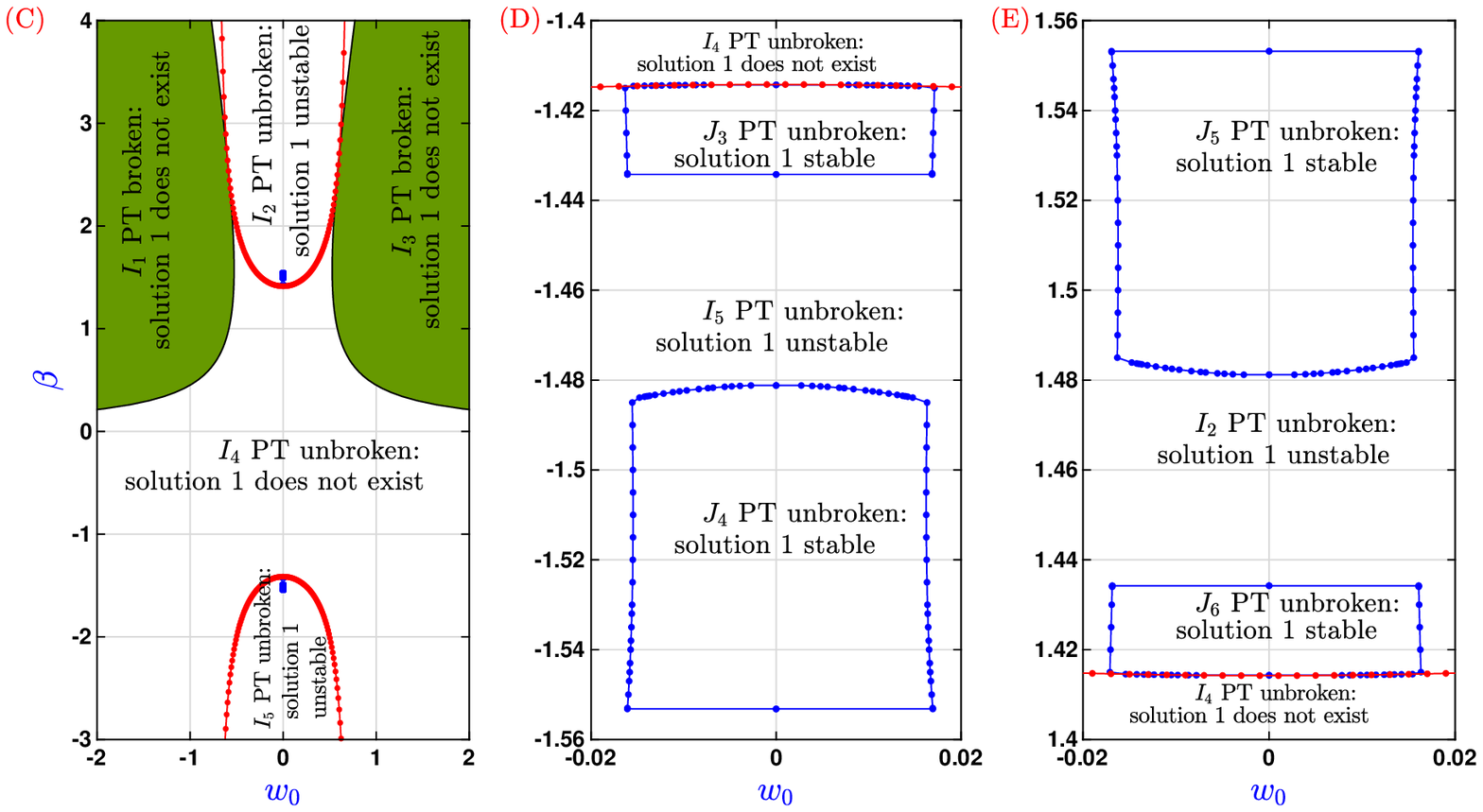}
	\caption{\label{Fig7.solution1g1phasetransition}(A) Phase transition of the linear operator (\ref{linearspectrum}) with potential (\ref{VWEx1}) and stable-unstable mode of the solution (\ref{solpsi0}) in $(w_0,v_0)$ plane for $g=1,\b=1$. (C) The phase transitions for the same linear operator and stable-unstable mode of the solution (\ref{solpsi0}) in $(w_0,\b)$ plane, for $g=1,v_0=0.5$.}
\end{figure}

We now consider two cases of $g$. In the first one, $g=-1$, the existence of the solution (\ref{solpsi0}) of the NLSE (\ref{nls}) in $(w_0,v_0)$ and $(w_0,\b)$ planes are shown in Fig. \ref{Fig6.solution1g-1phasetransition} (A) -(B). The space $(w_0,v_0)$ is divided into seven regions $I_i,~i=1-5,~J_1$ and $J_2$. In region $I_1$, solution (\ref{solpsi0}) does not exist. In addition, $\mathcal{PT}$ symmetric phase transition is shown in Fig. \ref{Fig6.solution1g-1phasetransition} (A). In green region $(I_5\cup J_1\cup J_2)$ linear operator $L$ (\ref{linearspectrum}) is $\mathcal{PT}$ broken and in white region $(\cup_{i=1}^4 I_i)$ $L$ is $\mathcal{PT}$ unbroken. The stable region for solution (\ref{solpsi0}) is shown in Fig. \ref{Fig6.solution1g-1phasetransition} (A) and it is bounded by blue and red curves, which is equal to $I_3\cup J_1\cup J_2$.  From this figure it is clear that $I_3$ is $\mathcal{PT}$ unbroken stable, whereas $J_1\cup J_2$ is $\mathcal{PT}$ broken stable, which is very small with respect to $I_3$ and total stable region $I_3\cup J_1\cup J_2$ is connected. The $(w_0,\b)$ space is divided into nine regions $(I_i,~i=1-9)$ and they are shown in Fig. \ref{Fig6.solution1g-1phasetransition} (B). From this figure one can see that, solution (\ref{solpsi0}) does not exists in $I_3\cup I_9$. The linear operator (\ref{linearspectrum}) is $\mathcal{PT}$ broken in green region $(I_1\cup I_5)$, whereas it is unbroken in white region $\left((\cup_{i=2}^4 I_i)\cup(\cup_{j=6}^9 I_j)\right)$. It is interesting to observe that solution (\ref{solpsi0}) is stable in a connected region $I_7$, which entirely lies in $\mathcal{PT}$ unbroken and there is no stable $\mathcal{PT}$ broken mode.  \cite{nath.optcom,nath.epjp}. 
In the second case $g=1$, the existence of BS in $(w_0,v_0)$ and $(w_0,\b)$ planes are shown in Fig. \ref{Fig7.solution1g1phasetransition} (A) and (C). From Fig. \ref{Fig7.solution1g1phasetransition} (A) we see that, BS does not exist below the red curve and the operator $L$ of (\ref{linearspectrum}) is broken in green region, which is entirely below the red curve. BS is stable in two disjoint regions $J_1$ and $J_2\subset I_1$ and they are shown in Fig. \ref{Fig7.solution1g1phasetransition} (B). From Fig. \ref{Fig7.solution1g1phasetransition} (C) we see that, BS exists in $I_2$ which is the upward interior of the red parabola and in $I_5$ which is the downward interior of red parabola. The stable regions exist in four small parts such as $J_3$, $J_4\subset I_2$ and $J_5$, $J_6\subset I_5$ and they are shown in Fig. \ref{Fig7.solution1g1phasetransition} (D) and (E). All parts are connected but $\ds\cup_{i=3}^{6} J_i$ is disconneted. For $g=1$, stable region in both planes are very small, which agree with \cite{23-42.33}. 
\subsection{Stability of DS}
\begin{figure}[h] 
	\centering
	\includegraphics[width=8cm,height=5cm]{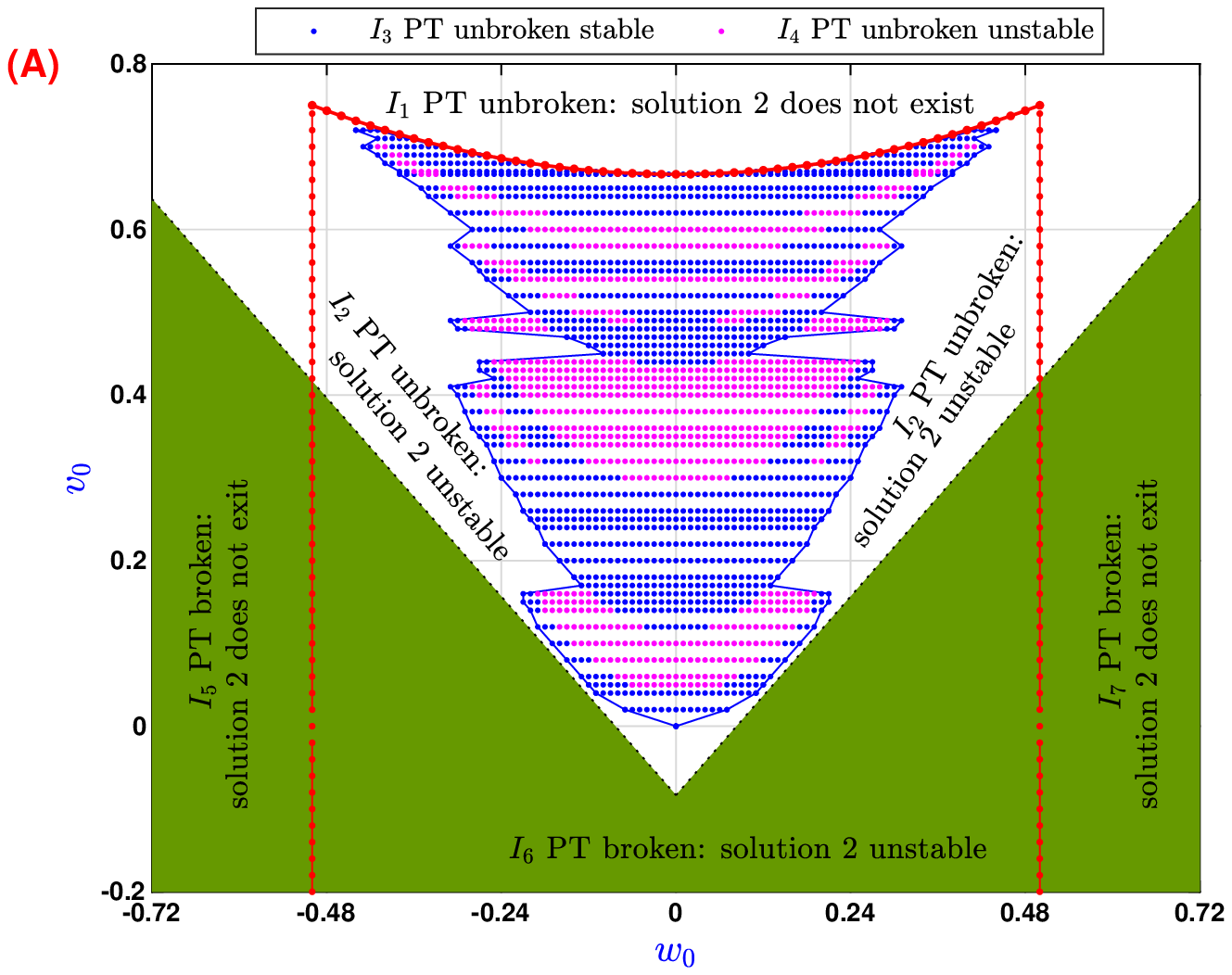}\\
	\includegraphics[width=8cm,height=5cm]{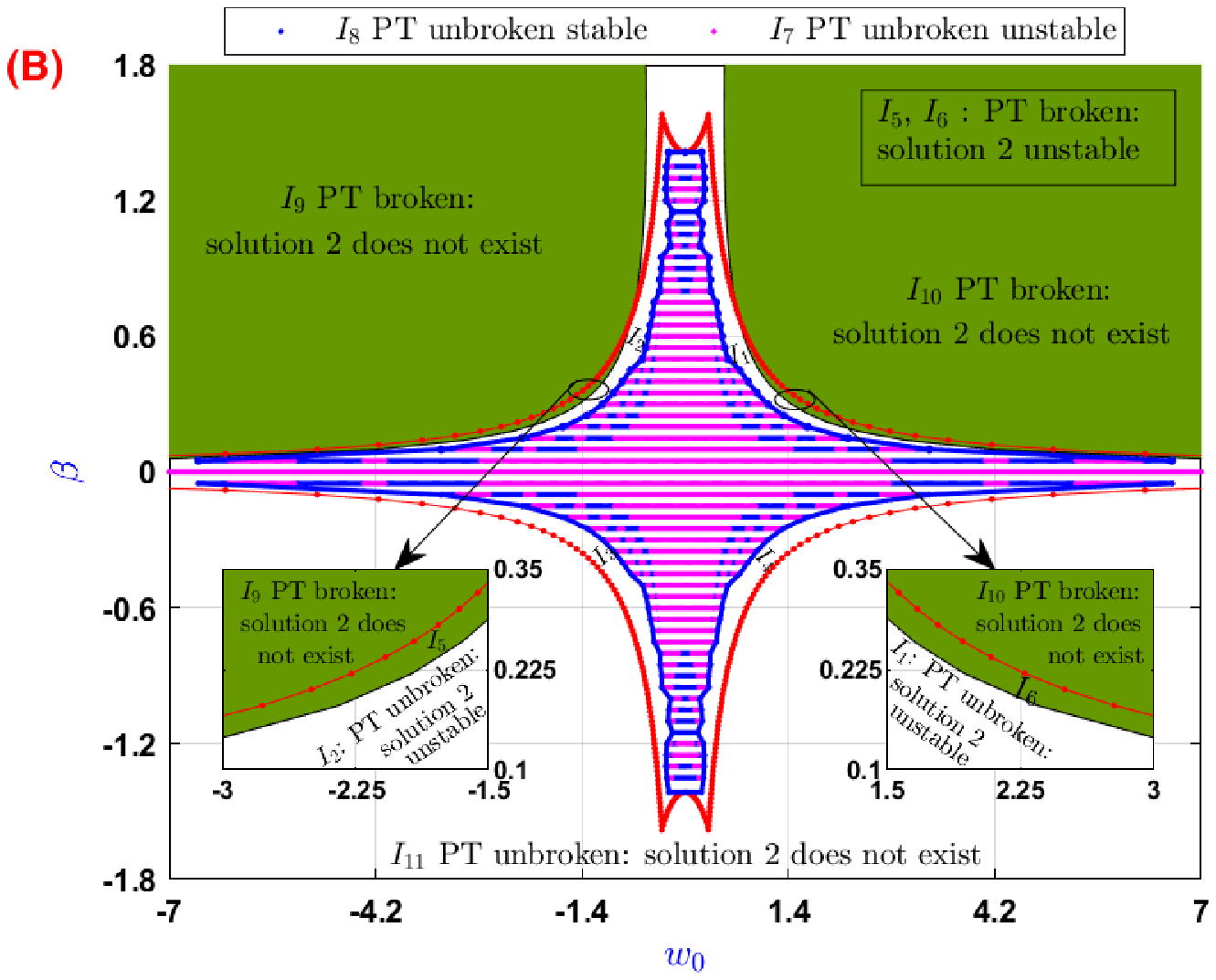}
	\caption{\label{Fig8.solution2g1phasetransition}(A) Phase transition of the linear operator $L$ (\ref{linearspectrum}) with potential (\ref{VWEx1}) and stable-unstable mode of the solution (\ref{solpsi1}) in $(w_0,v_0)$ plane, $g=1,\b=1,\d=1$. (B) Phase transition of the same linear operator and stable-unstable mode of the solution (\ref{solpsi1}) in $(w_0,\b)$ plane, for $g=1,v_0=0.5,\d=1$.}
\end{figure}

For $g=-1$ the DS is stable in small region in $(w_0,v_0)$ and $(w_0,\b)$ planes satisfying ${w_0}^2<\frac{1}{4\b^2}$ and some additional conditions. In particular, $\b=1,\d=1$, we have found a stable region in $(w_0,v_0)$ plane, such that $\frac{2+w_0^2}{3}\leq v_0\leq\frac{2+w_0^2}{3}+\epsilon$, where $0\leq\epsilon<0.1$. We now consider the second case $g=1$, DS exists in a domain which lies between two vertical lines $w_0^2< \f{1}{4\b^2}$ and a paraboic curve $v_0<\f{w_0^2\b^2+2}{\b^2+2}$ in $(w_0,v_0)$ plane. In particular, for $\b=1$, $\d=1$, the region is shown in Fig. \ref{Fig8.solution2g1phasetransition} (A) and the linear operator is broken in green region $\cup_{i=5}^{7} I_i$. Stable modes of DS are marked by \emph{blue dots, $I_3$} and \emph{magenta dots, $I_4$} represent unstable modes. It is to be found that, stable modes are exist within $\mathcal{PT}$ unbroken region. We did not find stable mode in broken region after several trials. Similarly, existence of (\ref{solpsi1}) and $\mathcal{PT}$ broken-unbroken phase transition shown in Fig. \ref{Fig8.solution2g1phasetransition} (B) in $(w_0,\b)$ plane, for $v_0=0.5,\d=1$. The space $(w_0,\b)$ is divided into eleven regions $I_i,~i=1-11$. The linear operator $L$ is broken in $I_5\cup I_6\cup I_9\cup I_{10}$ and unbroken in $\ds\cup_{i=1}^{4} I_i\cup I_7\cup I_8\cup I_{11}$ and DS exists in a region bounded by red curves. Stable modes of DS are marked by \emph{blue dots, $I_8$} and unstable in $\ds\cup_{i=1}^{7}I_i$. The DS has stable modes only in $\mathcal{PT}$ unbroken region. To confirm all these stable modes of BS and DS, we will investigate their excitations in the next section \ref{sec4.excitation}.

\section{Excitation}\label{sec4.excitation}
Now, we will focus on nature of solutions of time dependent NLSE with complex $\mathcal{PT}$-symmetric potential
\beq\label{nlstExcitation}
i\,\Psi_t=-\Psi_{xx}+\left(V(x,t)+i\,W(x,t)\right)\Psi+g|\Psi|^2\Psi,
\eeq
where $V(x,t), W(x,t)$ are given by Eq.(\ref{VWEx1}) with $v_0\rightarrow v_0(t)$ and $w_0\rightarrow w_0(t)$ satisfy \cite{excitation}
\beq\label{v0w0betaExcitation}
\left\{v_0,w_0,\b\right\}(t)=\left\{\ba{ll}\left(\left\{v_{02},w_{02},\b_{02}\right\}-\left\{v_{01},w_{01},\b_{01}\right\}\right)f(t)\\ +\left\{v_{01},w_{01},\b_{01}\right\},0\le t\le 5000\\
\left\{v_{02},w_{02},\b_{02}\right\},  t>5000\ea\right.,
\eeq
where $f$ is a real valued function of $t$. It is easy to verify that, solutions $(\ref{solpsi0}) and (\ref{solpsi1})$ do not satisfy Eq.(\ref{nlstExcitation}) but they satisfy for initial time $t=0$ and $t\geq5000$. In this paper we have considered $f(t)=\sin\f{\pi t}{10000}$, $0\le t\le 5000$.
\begin{figure}[h]
	\centering
	\includegraphics[width=5cm,height=3cm]{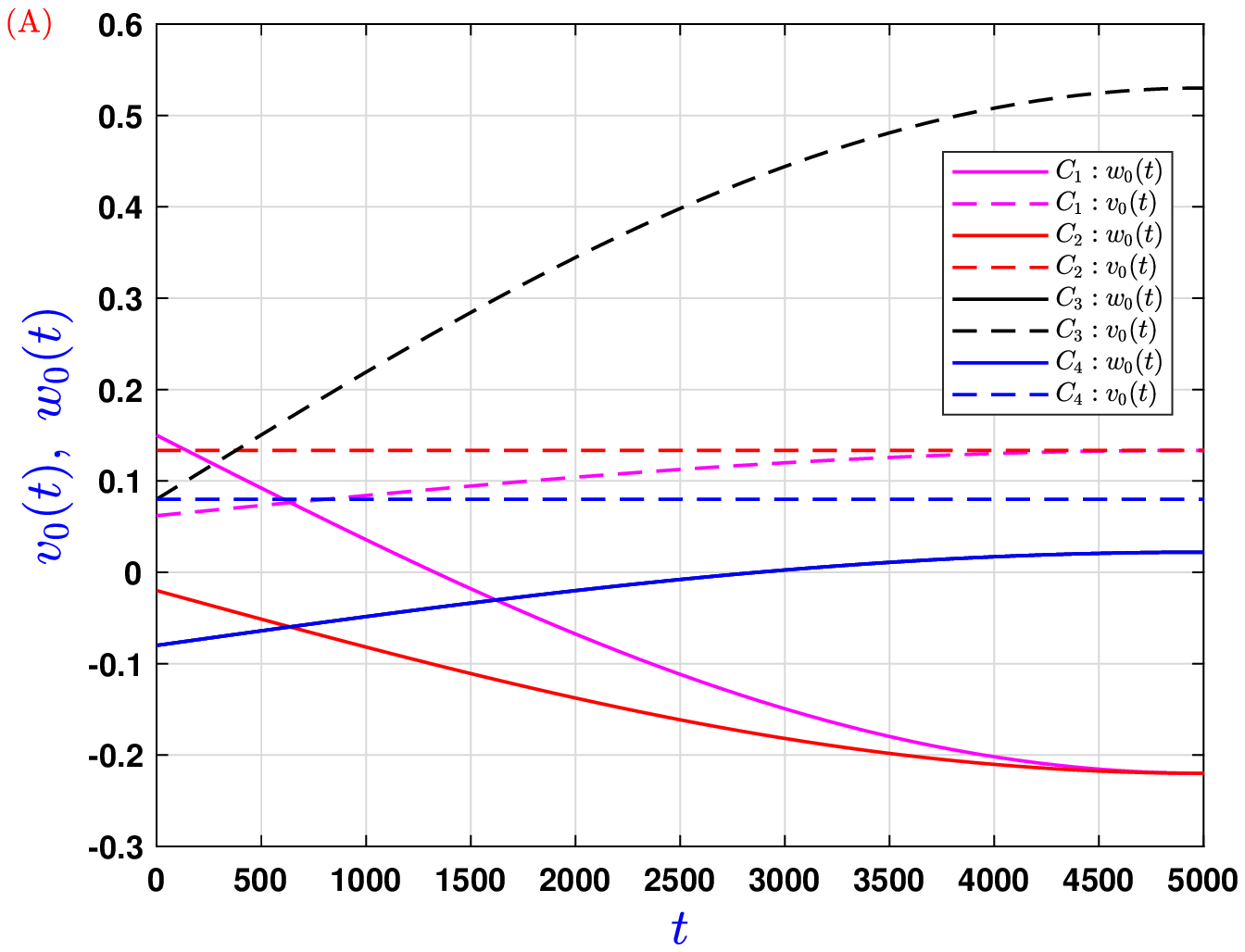}\includegraphics[width=5cm,height=3cm]{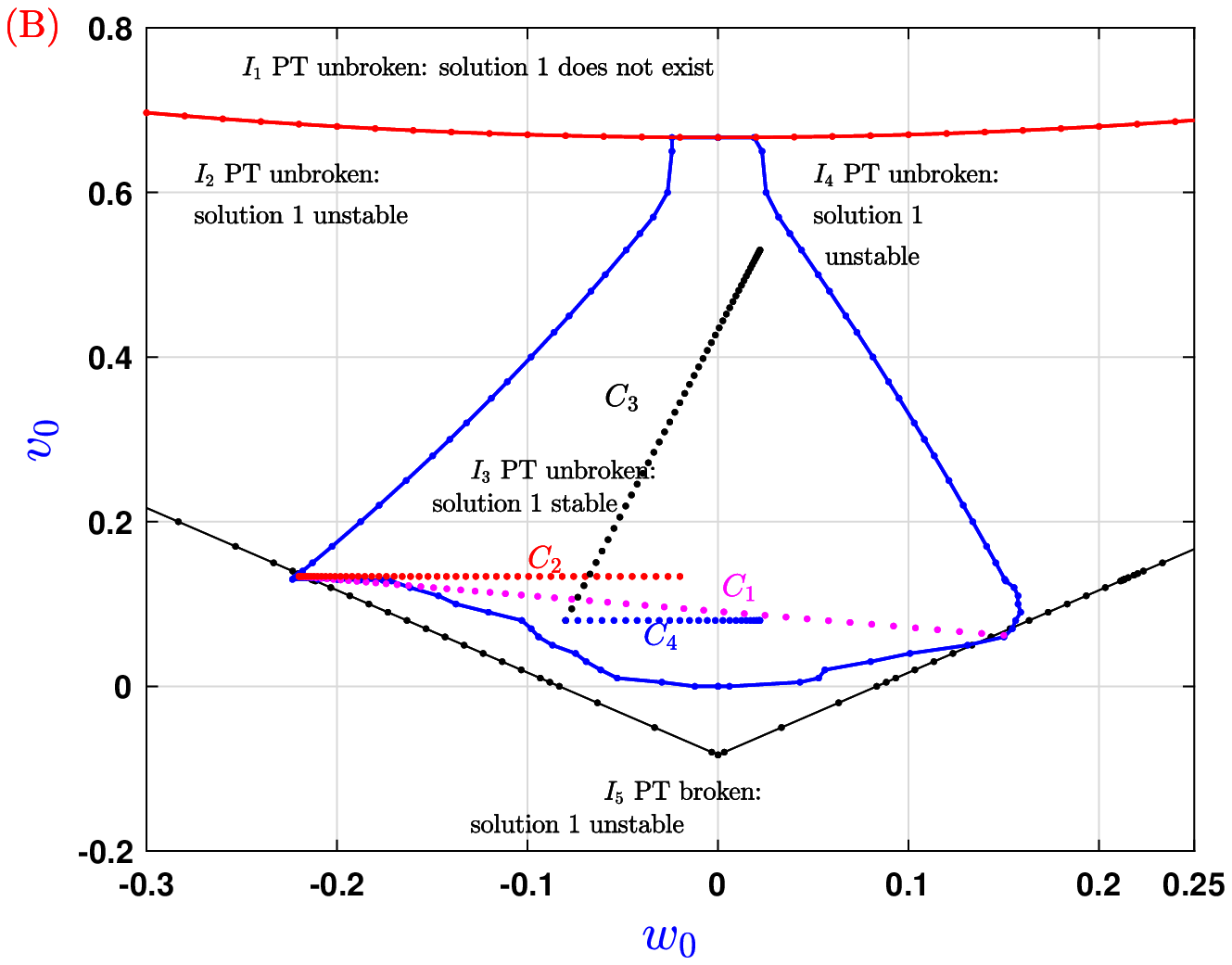}\\
	\includegraphics[width=5cm,height=3cm]{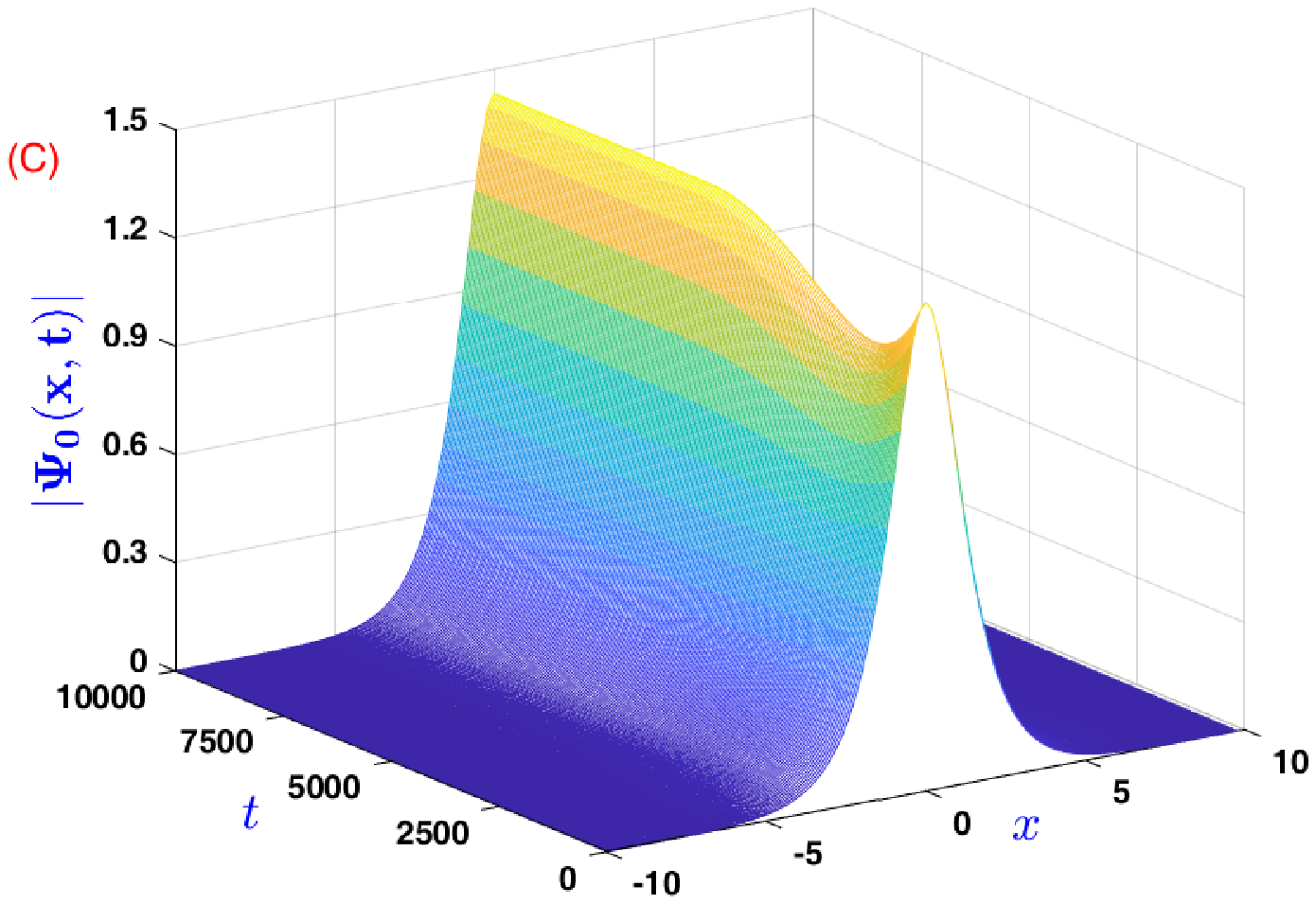}\includegraphics[width=5cm,height=3cm]{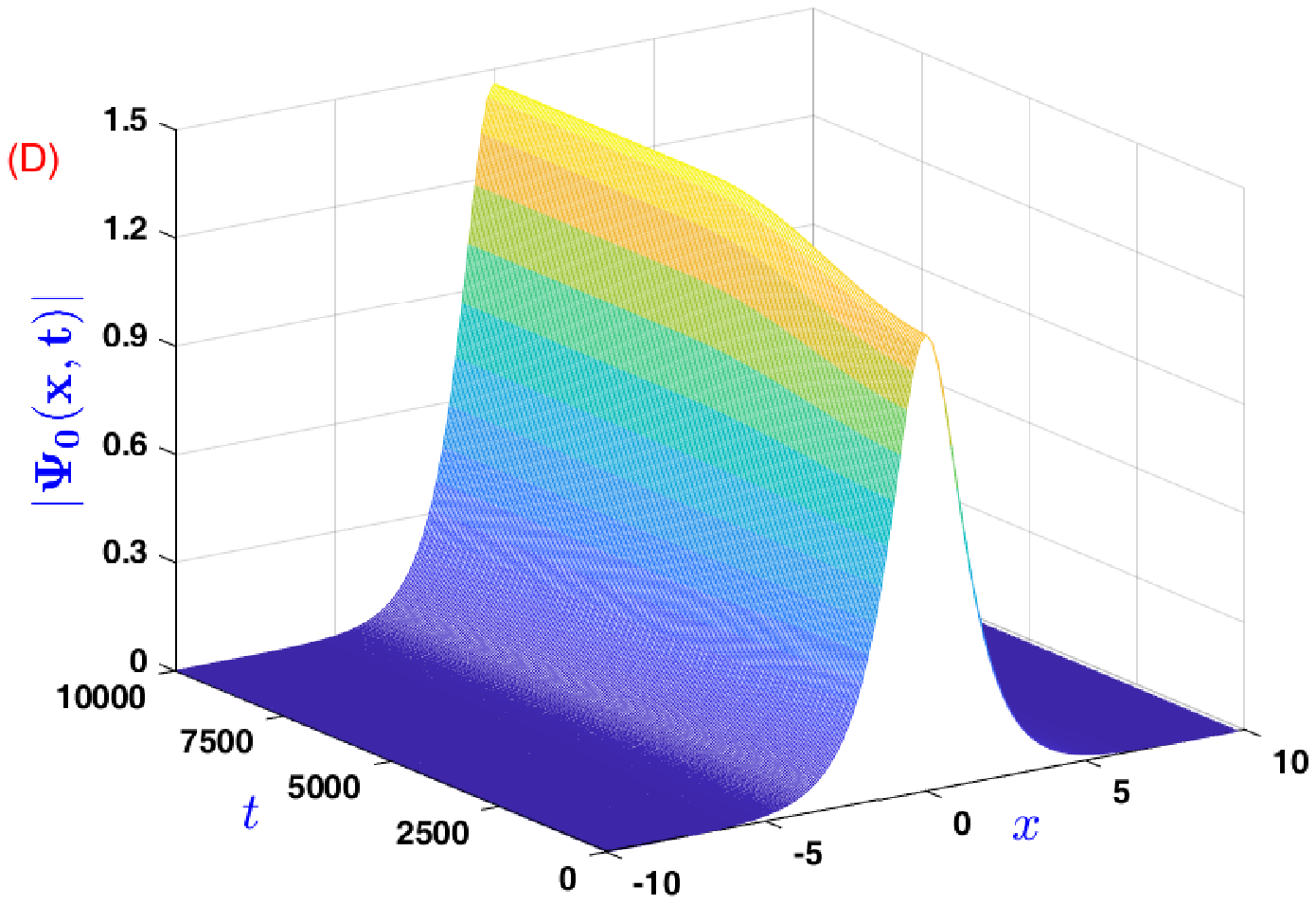}\\
	\includegraphics[width=5cm,height=3cm]{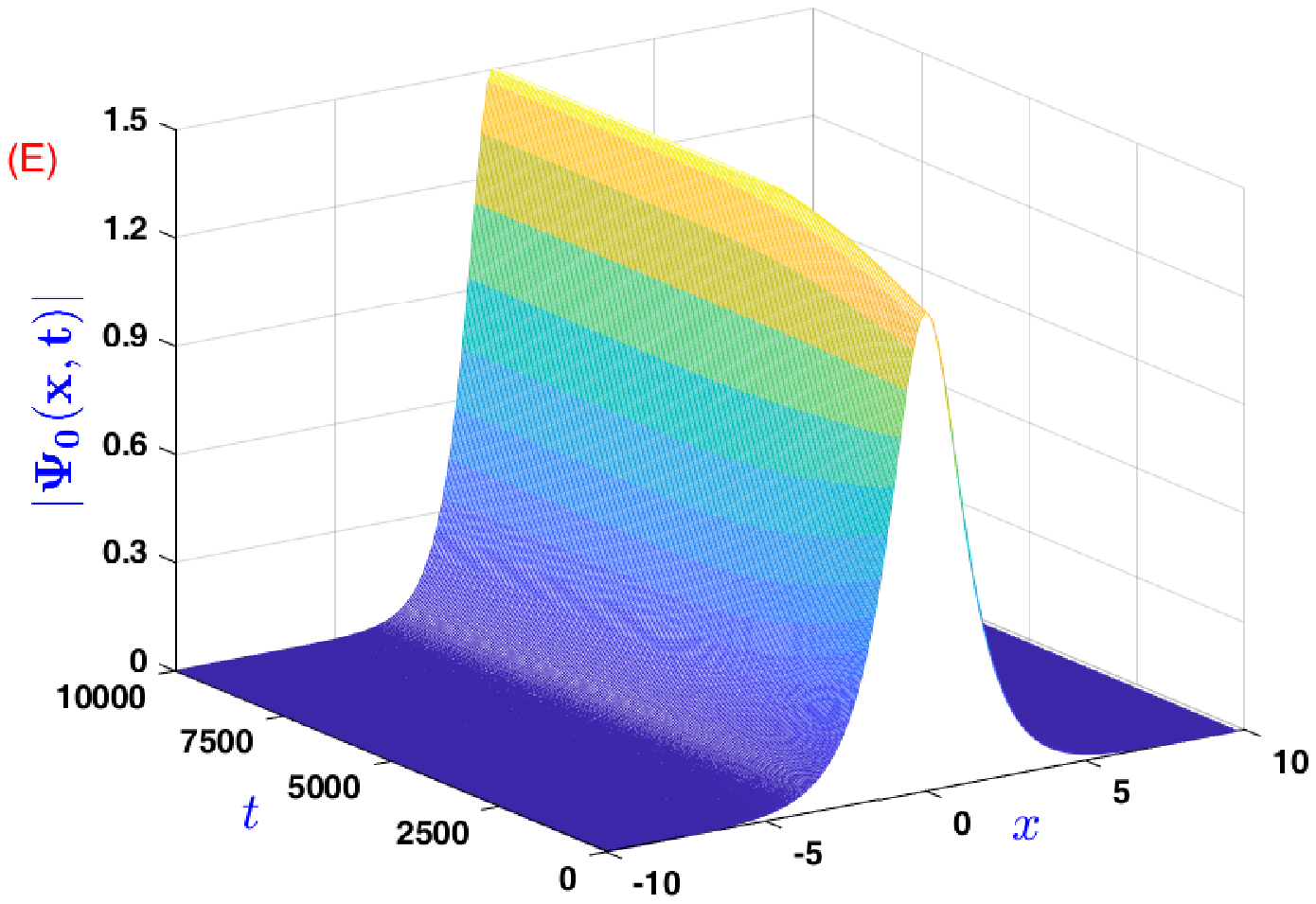}\includegraphics[width=5cm,height=3cm]{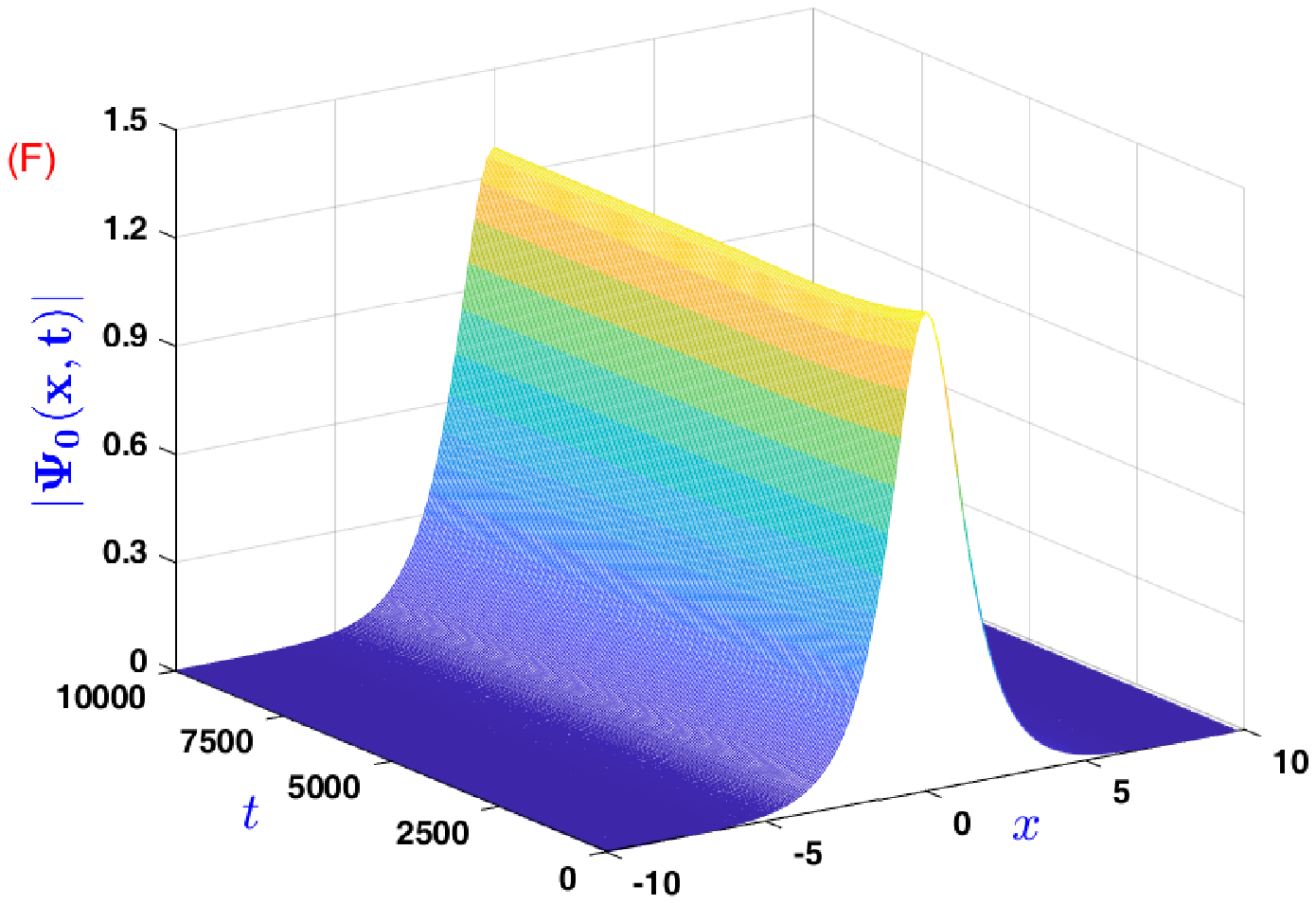}~
	\caption{\label{Fig9.solution1g-1excitationw0v0} (A) Shows time dependent potential amplitudes $w_0, v_0$ for different curves. (B) Shows that stable and unstable regions of BS and four curves $C_i,~i=1-4$. Excitations of stable nonlinear localized state of (\ref{nlstExcitation}) with intial condition (\ref{solpsi0}), for $g=-1$, $\b=1$,
		(C) $v_{01}=0.062,v_{02}=0.1335,w_{01}=0.15,w_{02}=-0.22$ (D)$v_{01}=0.1335,v_{02}=0.1335,w_{01}=-0.02,w_{02}=-0.22$
		(E) $v_{01}=0.08,v_{02}=0.53,w_{01}=-0.08,w_{02}=0.022$   (F) $v_{01}=0.08,v_{02}=0.08,w_{01}=-0.08,w_{02}=0.022$.}
\end{figure}
\subsection{Excitation of BS}
We have displayed excitations of BS of (\ref{nlstExcitation}), with initial condition (\ref{solpsi0}) in Fig. \ref{Fig9.solution1g-1excitationw0v0} (C) -(F), for $g=-1$ along four curves $C_1$ - $C_4$. Each curve has initial point $P_1(w_{01},v_{01})$ and final point $P_2(w_{02},v_{02})$. The time dependent potential amplitudes $ \min\{v_{01},v_{02} \}\le v_0(t)\le \max\{v_{01},v_{02}\}$ and $\min\{w_{01},w_{02} \}\le w_0(t)\le \max\{w_{01},w_{02} \}$ satisfy the Eq.(\ref{v0w0betaExcitation}) and they are plotted in Fig. \ref{Fig9.solution1g-1excitationw0v0} (A) with respect to $t$ and the corresponding locus of $(w_0(t),v_0(t))$ for $0\le t\le 5000$, represent curves $C_j~,j=1,2,3,4$ and they are shown in Fig. \ref{Fig9.solution1g-1excitationw0v0} (B). 
The curve $C_1:P_2(-0.22,0.1335)\nwarrow P_1(0.15,0.062)$ is started from $P_1(\in J_2,~\mbox{broken~region})$ to $P_2(\in J_1,~\mbox{broken~region})$ via unbroken region $(I_3)$, whereas $C_2:P_2(-0.22,0.1335)\longleftarrow P_1(-0.02,0.1335)$ is started from $P_1(\in I_3,~\mbox{unbroken~region})$ to $P_2(\in J_1,~\mbox{broken~region})$. Therefore, excitation Fig. \ref{Fig9.solution1g-1excitationw0v0} (C) started from a broken point passes through unbroken region and then eneded to a broken point, whereas, excitation Fig. \ref{Fig9.solution1g-1excitationw0v0} (D) started from an unbroken point and then going to broken region. 
On the otherhand Fig. \ref{Fig9.solution1g-1excitationw0v0} (E) and (F) are excitations along $C_3:P_1(-0.08,0.08)\nearrow P_2(0.022,0.53)$ and $C_4:P_1(-0.08,0.08)\longrightarrow P_2(0.022,0.08)$ and they lie within unbroken region $I_3$. One can see that, curves $C_1$, $C_3$ are look like oblique whereas, $C_2$, $C_4$ are like horizontal but their directions are different and all excitations in $(w_0,v_0)$ plane are stable. We have checked that, BS is stable for all $(w_0,v_0)\in I_3\cup J_1\cup J_2$ and its excitation is stable along any curve in any direction which lies in $I_3\cup J_1\cup J_2$. 
\begin{figure}[h]
	\centering
	\includegraphics[width=5cm,height=3cm]{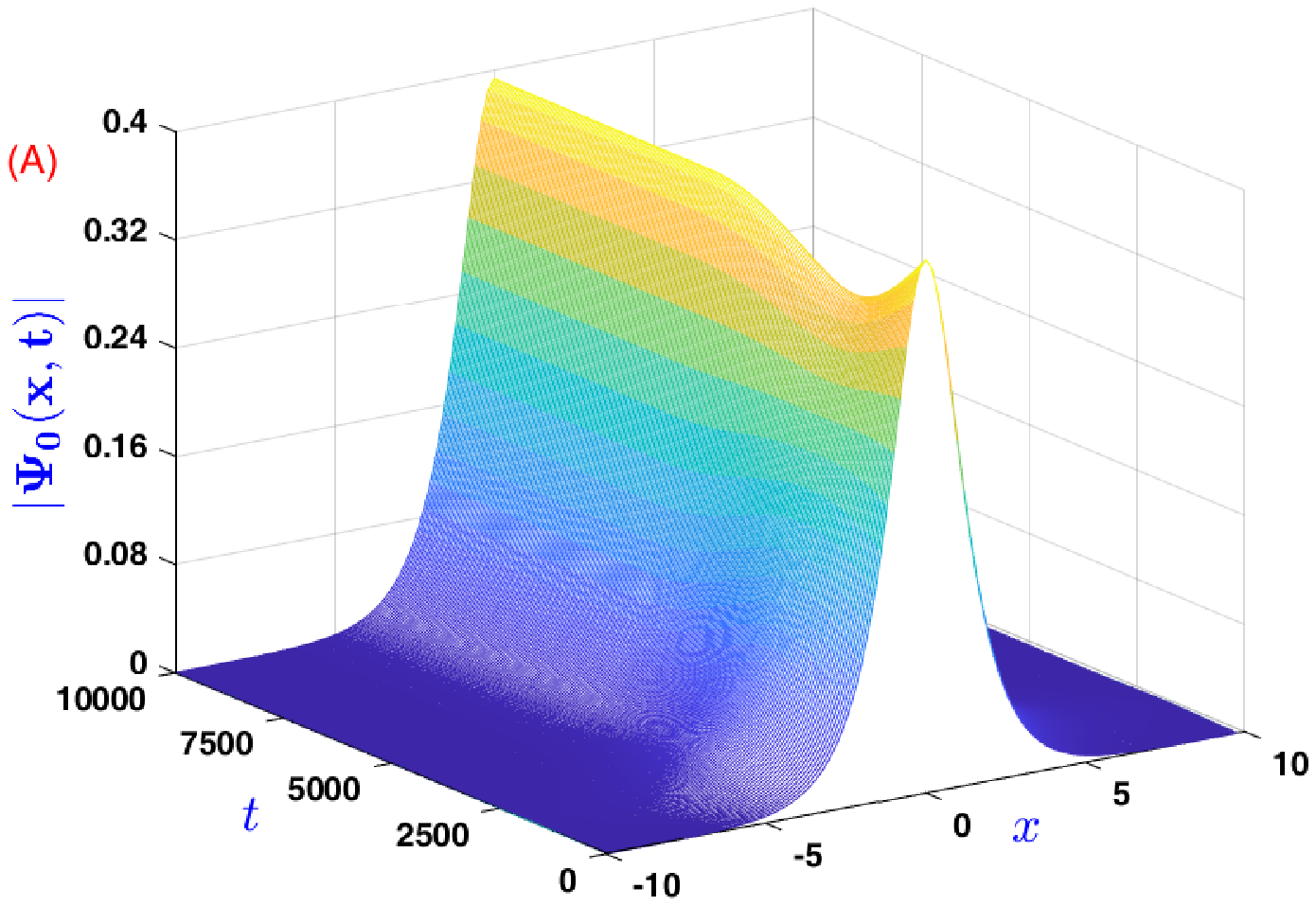}\includegraphics[width=5cm,height=3cm]{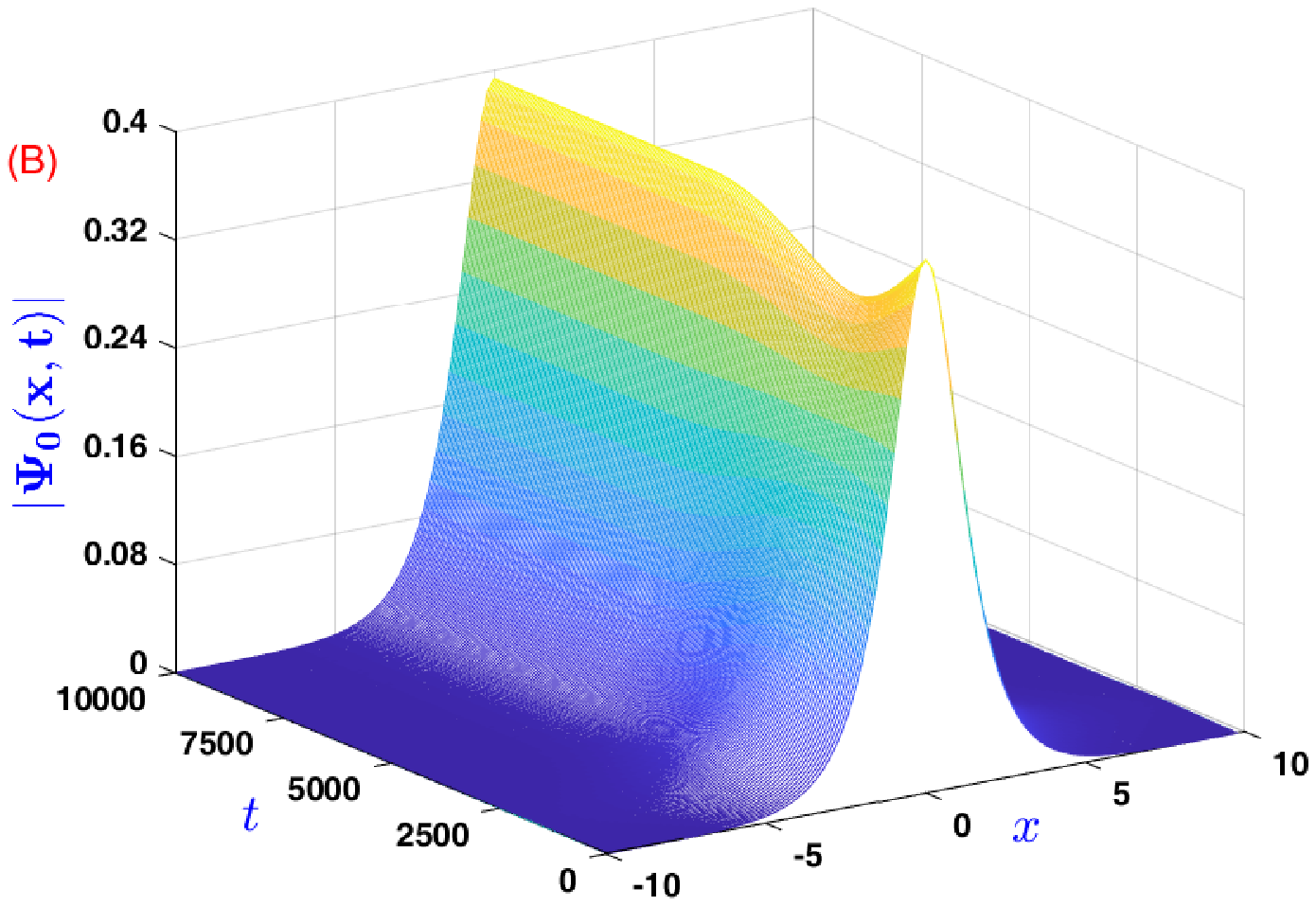}\\
	\includegraphics[width=5cm,height=3cm]{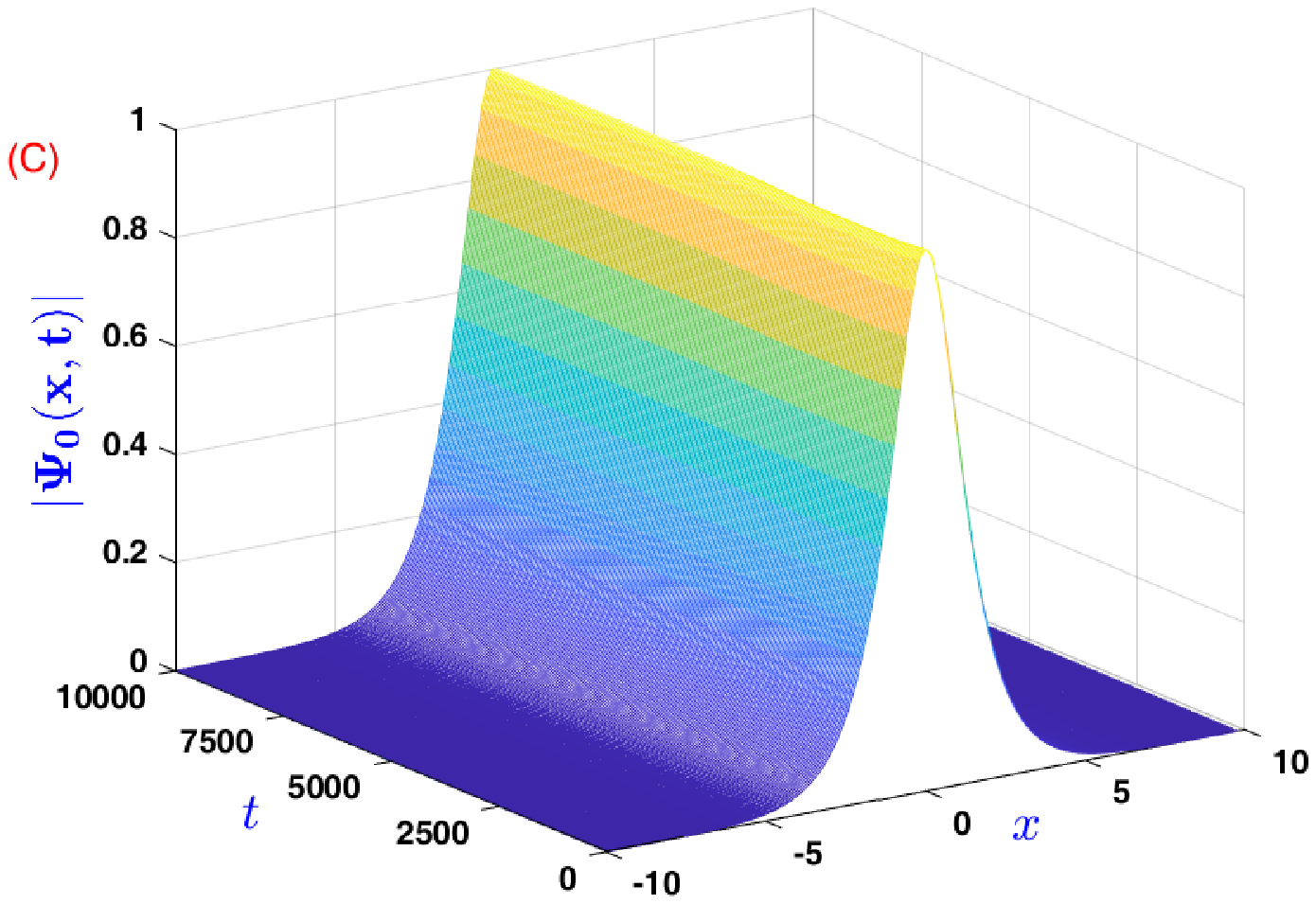}\includegraphics[width=5cm,height=3cm]{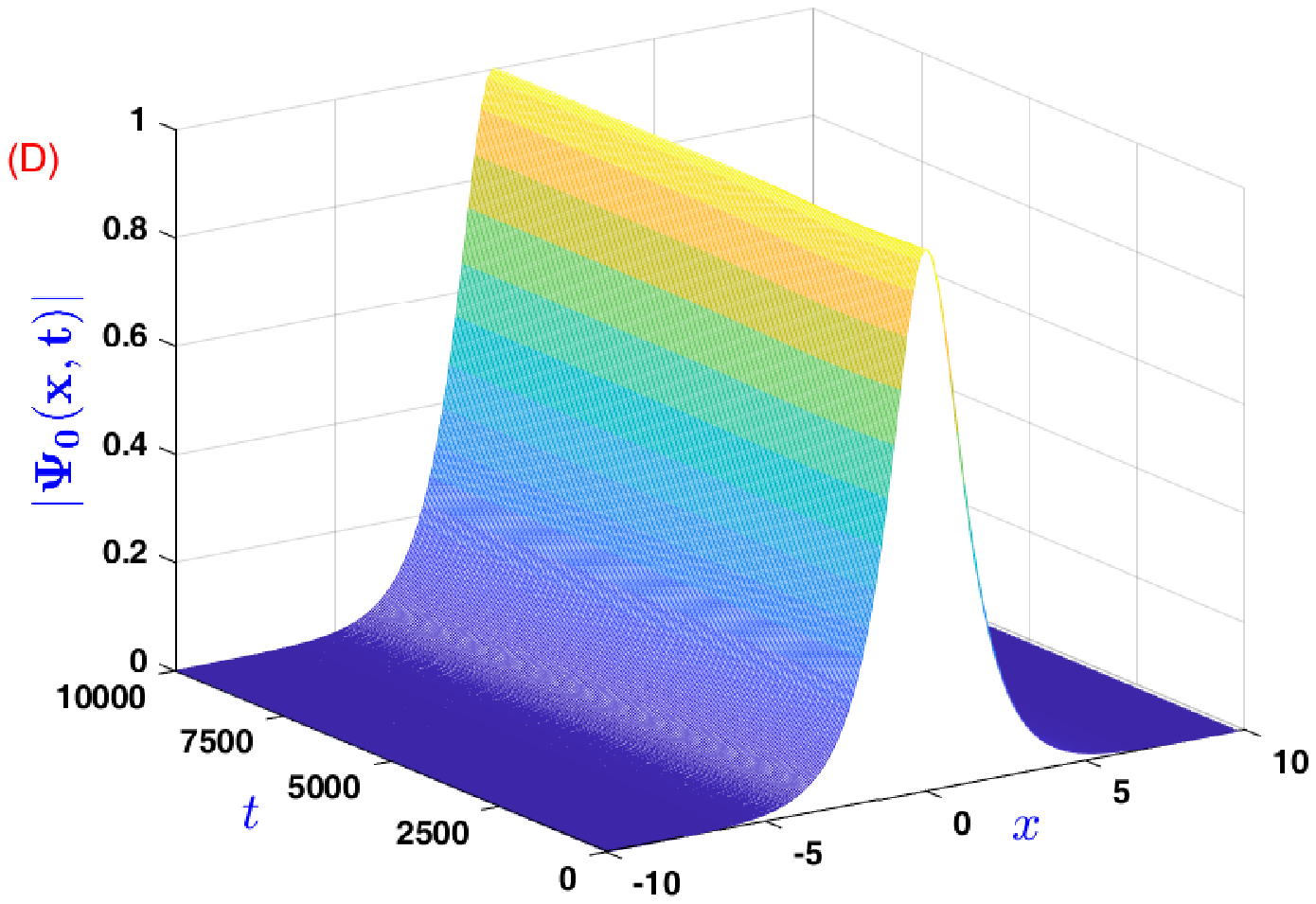}
	\caption{\label{Fig10.solution1g-1excitationw0beta} Excitations of stable nonlinear localized state of (\ref{nlstExcitation}) with initial condition (\ref{solpsi0}), for $g=-1$, $v_0=0.5$,
		(A) $\b_{01}=-1.30,\b_{02}=1.30,w_{01}=-0.017,w_{02}=0.017$ (B) $\b_{01}=-1.30,\b_{02}=1.30,w_{01}=-0.017,w_{02}=-0.017$
		(C) $\b_{01}=-0.02,\b_{02}=0.02,w_{01}=-4.7,w_{02}=4.7$     (D) $\b_{01}=-0.02,\b_{02}=-0.02,w_{01}=-4.7,w_{02}=4.7$.}
\end{figure}
Similarly, in $(w_0,\b)$ plane for $g=-1$, $v_0=0.5$ excitations of (\ref{nlstExcitation}) with initial solution (\ref{solpsi0}) are shown in Figs. \ref{Fig10.solution1g-1excitationw0beta} (A) -(D) along four curves $(C_5: P_1(-0.017,-1.3)\nearrow P_2(0.017,1.3);~ C_6:P_1(-0.017,-1.3)\uparrow P_2(-0.017,1.3);~ C_7:P_1(-4.7,-0.02)\nearrow P_2(4.7,0.02);~ C_8:P_1(-4.7,-0.02)\rightarrow P_2(4.7,-0.02))$. From Figs. \ref{Fig10.solution1g-1excitationw0beta} (A) -(D), one can see that BS and its excitations are stable. We observe that, excitation of BS of (\ref{nlstExcitation}) with intial state (\ref{solpsi0}) is stable along any curve in any direction which lies in $\mathcal{PT}$ unbroken region $I_7$ of Fig. \ref{Fig6.solution1g-1phasetransition} (B) in $(w_0,\b_0)$ plane. 
\begin{figure}[h]
	\centering
	\includegraphics[width=5cm,height=3cm]{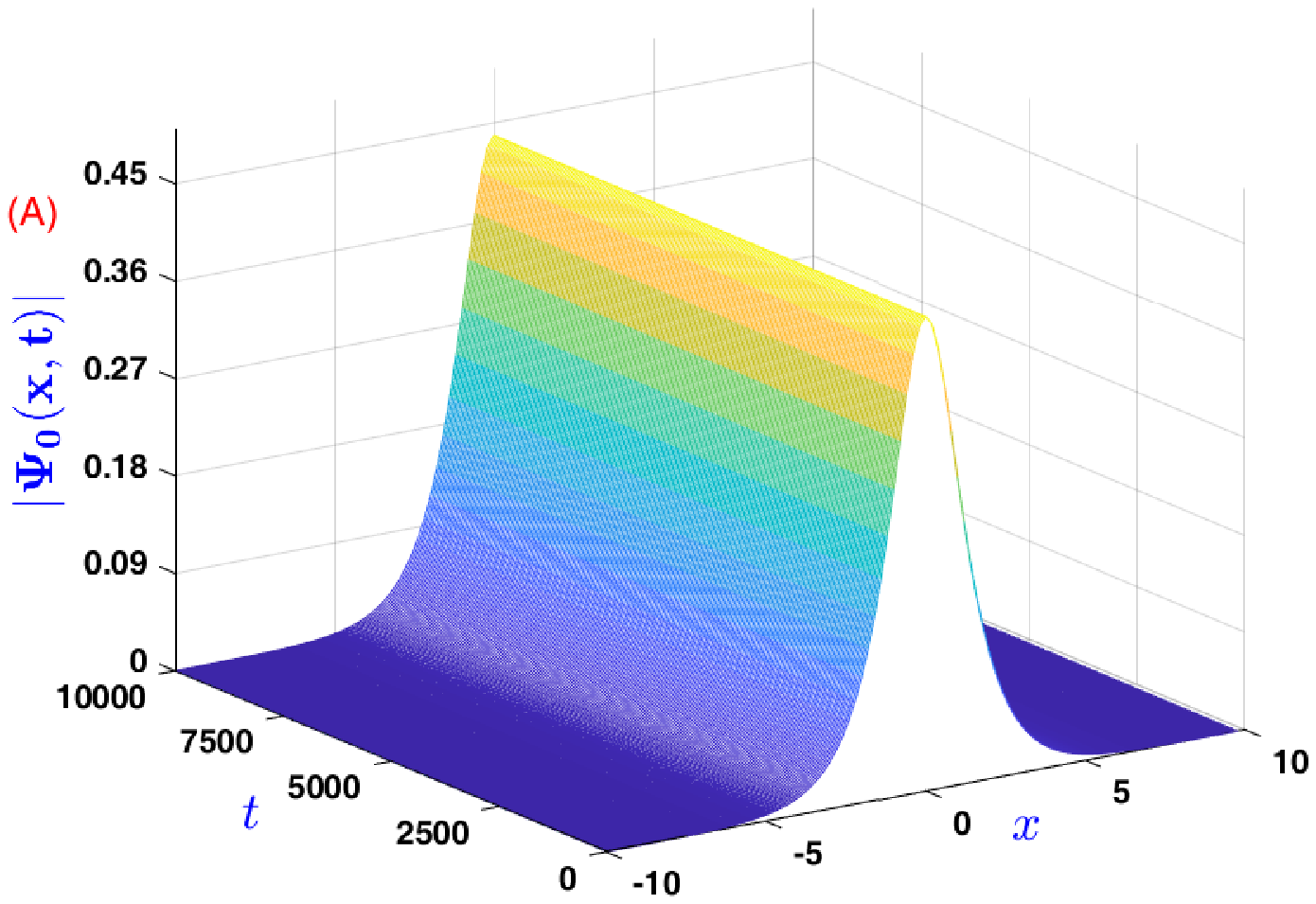}\includegraphics[width=5cm,height=3cm]{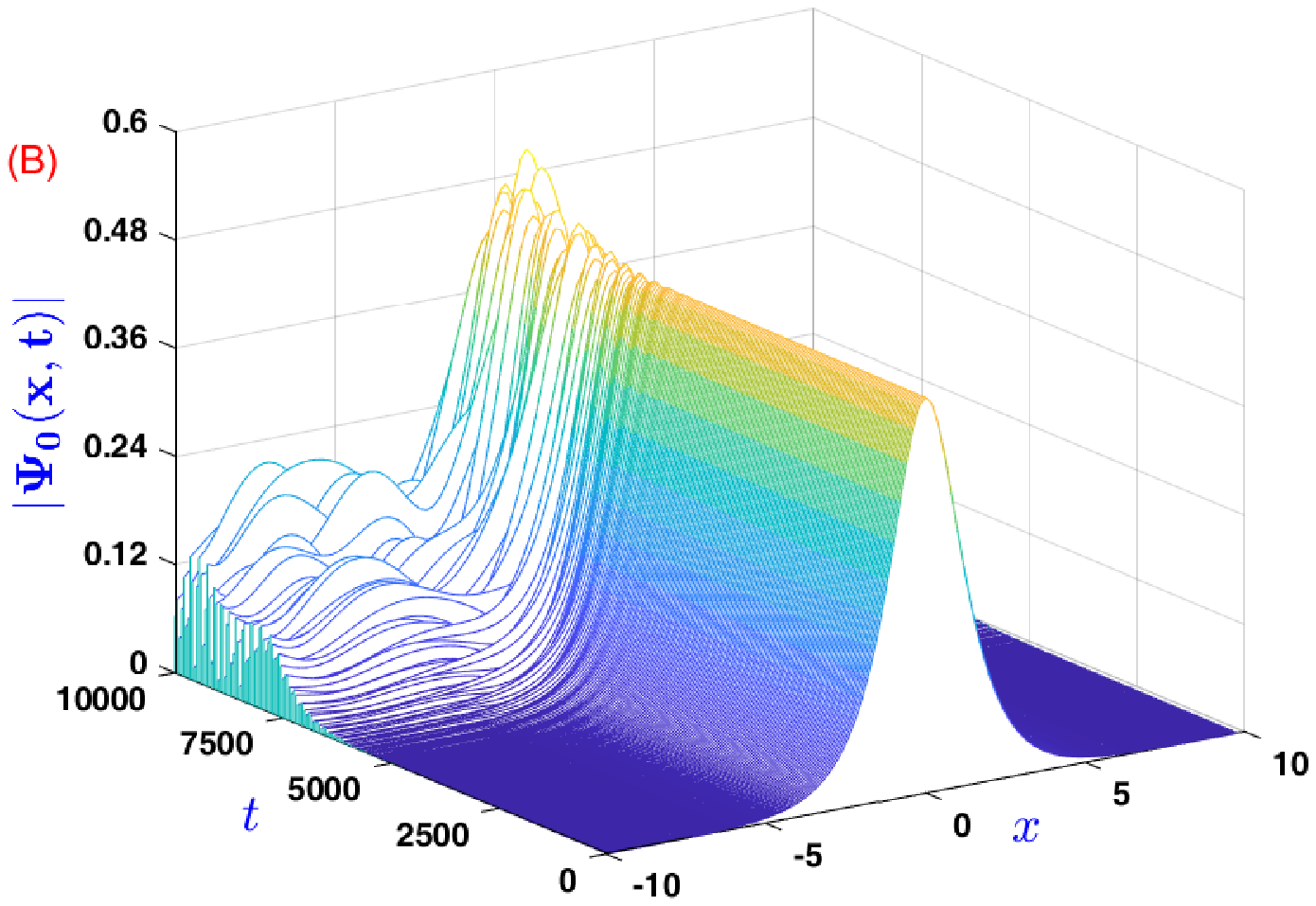}
	\caption{\label{Fig11} Excitations of nonlinear localized state of (\ref{nlstExcitation}) with initial condition (\ref{solpsi0}) in $(w_0,v_0)$ plane for $g=1$, $\b=1$, (A) $v_{01}=0.731, v_{02}=0.731, w_{01}=0.024, w_{02}=-0.024$; (B) $v_{01}=0.731, v_{02}=0.731, w_{01}=0.024, w_{02}=0.076$.}
\end{figure}
\begin{figure}[h]
	\centering
	\includegraphics[width=5cm,height=3cm]{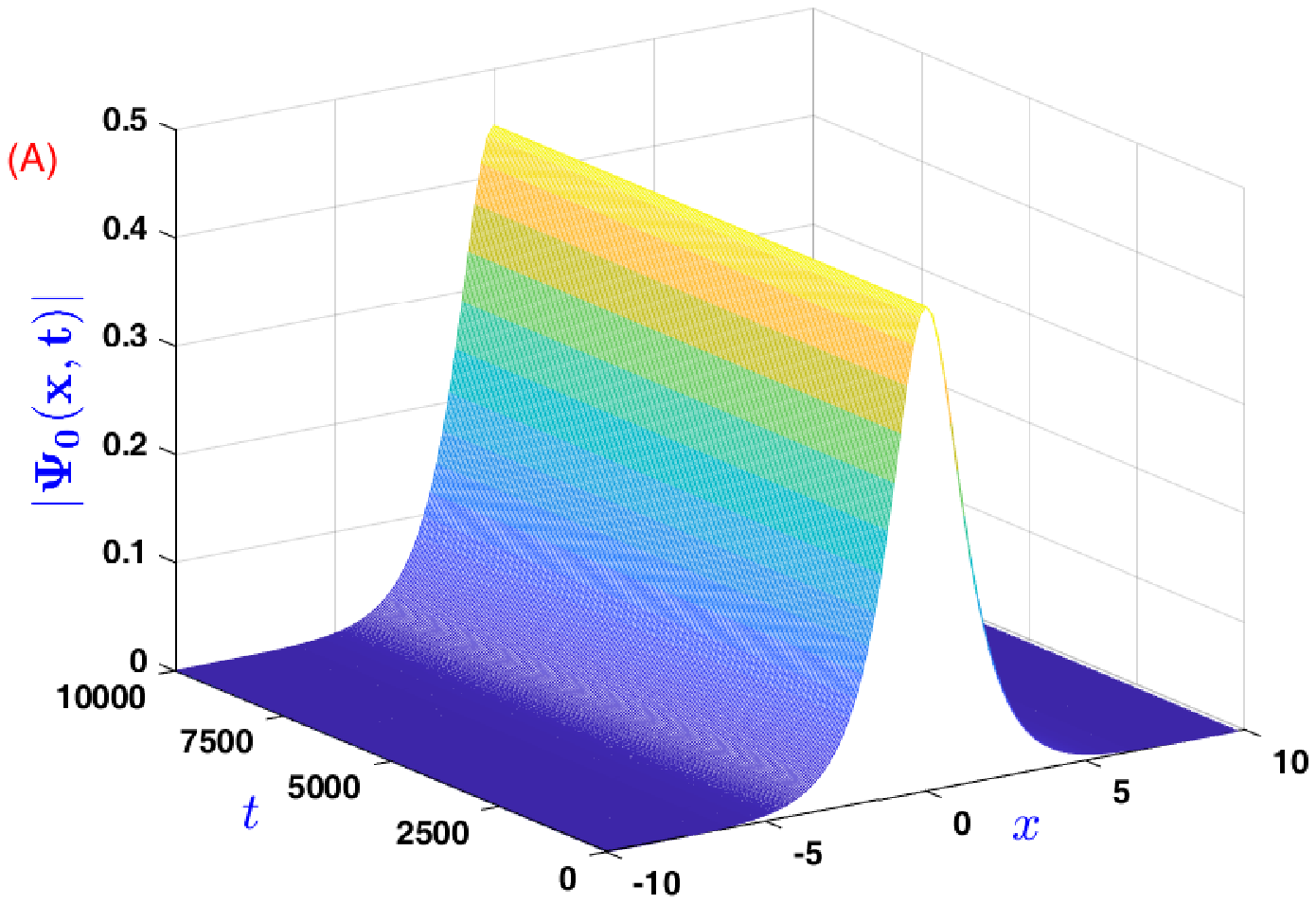}\includegraphics[width=5cm,height=3cm]{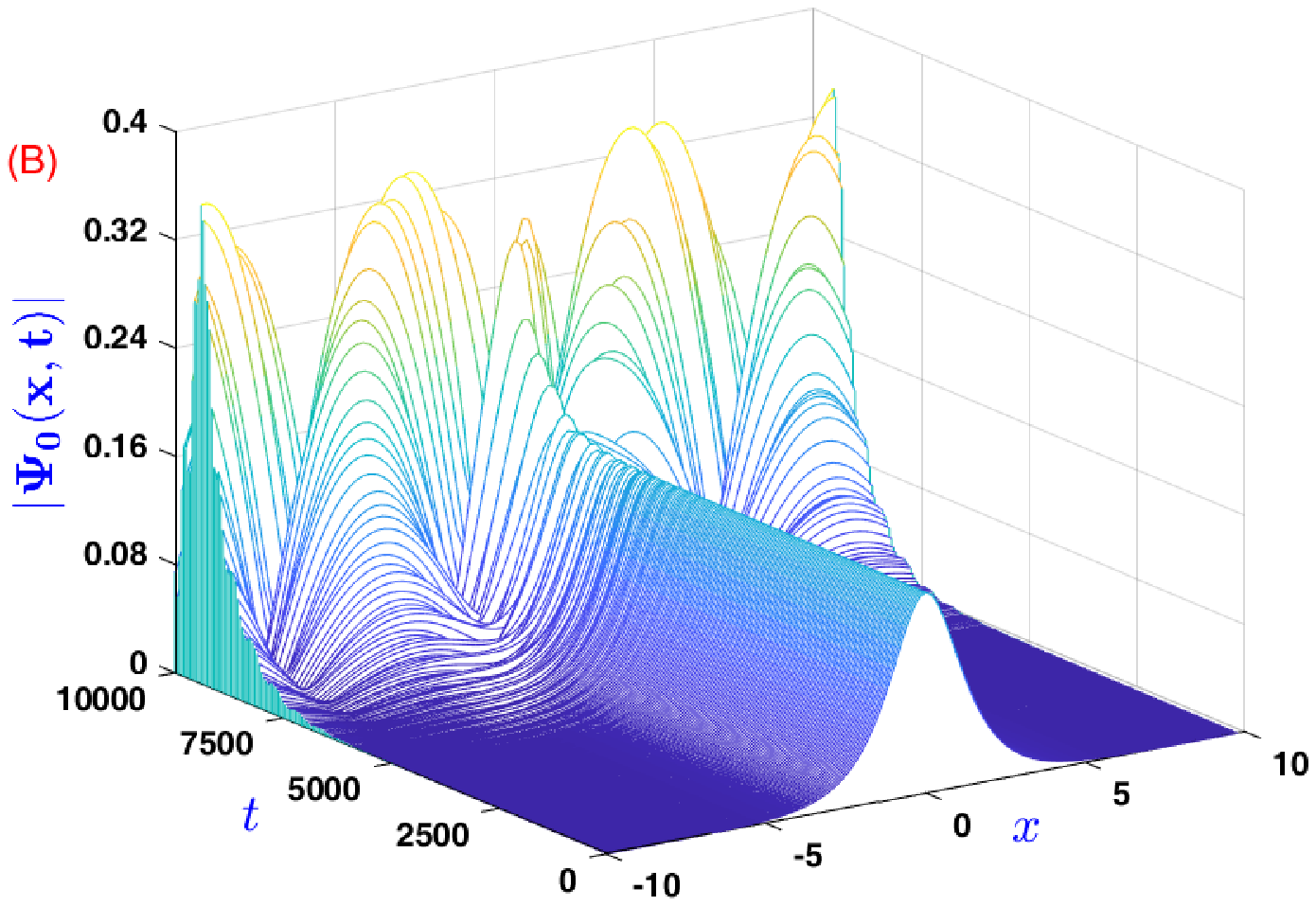}
	\caption{\label{Fig12.solution1g1excitationw0beta} Excitations of stable nonlinear localized state of (\ref{nlstExcitation}) with initila condition (\ref{solpsi0}) in $(w_0,\b)$ plane for $g=1$, $v_0=0.5$, (A) $\b_{01}=-1.550,\b_{02}=-1.550,w_{01}=-0.016,w_{02}=0.016$; (B) $w_{01}=-0.015,\b_{01}=1.43, w_{02}=0.017, \b_{02}=1.44$}
\end{figure}
Similarly, in $(w_0,v_0)$ for $g=1$, excitations of BS with same initial condition (\ref{solpsi0}) along $C_9: P_2 (-0.024,0.731)\leftarrow P_1(0.024,0.731)~(\in J_1)$ and $C_{10}: P_1(0.024,0.731)~(\in J_1) \rightarrow P_2 (0.076,0.731)$ are shown in Fig. \ref{Fig11} (A) and (B) respectively. The curve $C_9$ lies in the stable region $J_1$ of Fig. \ref{Fig7.solution1g1phasetransition} (B) and the curve $C_{10}$ is started from a stable point of $J_1$ and to an unstable point of $I_1$ of Fig. \ref{Fig7.solution1g1phasetransition} (B). For $C_9$ BS is stable and for $C_{10}$ it is unstable and we observe that, BS is stable for all $(w_0,v_0)\in J_1\cup J_2$, where $J_1$ and $J_2$ are connected regions but $J_1\cup J_2$ is disconnected.

Similarly, in $(w_0,\b)$ plane for $g=1, $ $(w_0,\b)$ plane excitations of BS of (\ref{nlstExcitation}) along $C_{11}: P_1(-0.016,-1.55)~(\in J_4) \rightarrow P_2(0.016,-1.55)~(\in J_4)$ and $C_{12}:P_1(-0.015,1.43)\in J_6\nearrow P_2(0.017,1.44)\in I_2$ curves are shown in Fig. \ref{Fig12.solution1g1excitationw0beta} (A) and (B) respectively. Here, $C_{11}$ lies in a stable region $J_4$ of Fig. \ref{Fig7.solution1g1phasetransition} (D) and for $C_{12}$, some part lies in a stable region $J_6$ and the remaining part lies in unstable region $I_2$ of Fig. \ref{Fig7.solution1g1phasetransition} (E). Once can sees that, BS is stable for $C_{11}$ and it is unstable for $C_{12}$. Similarly, BS is stable in $\cup_{i=3}^6 J_i$.

\subsection{Excitation of DS}
\begin{figure}[h]
	\centering
	\includegraphics[width=5cm,height=3cm]{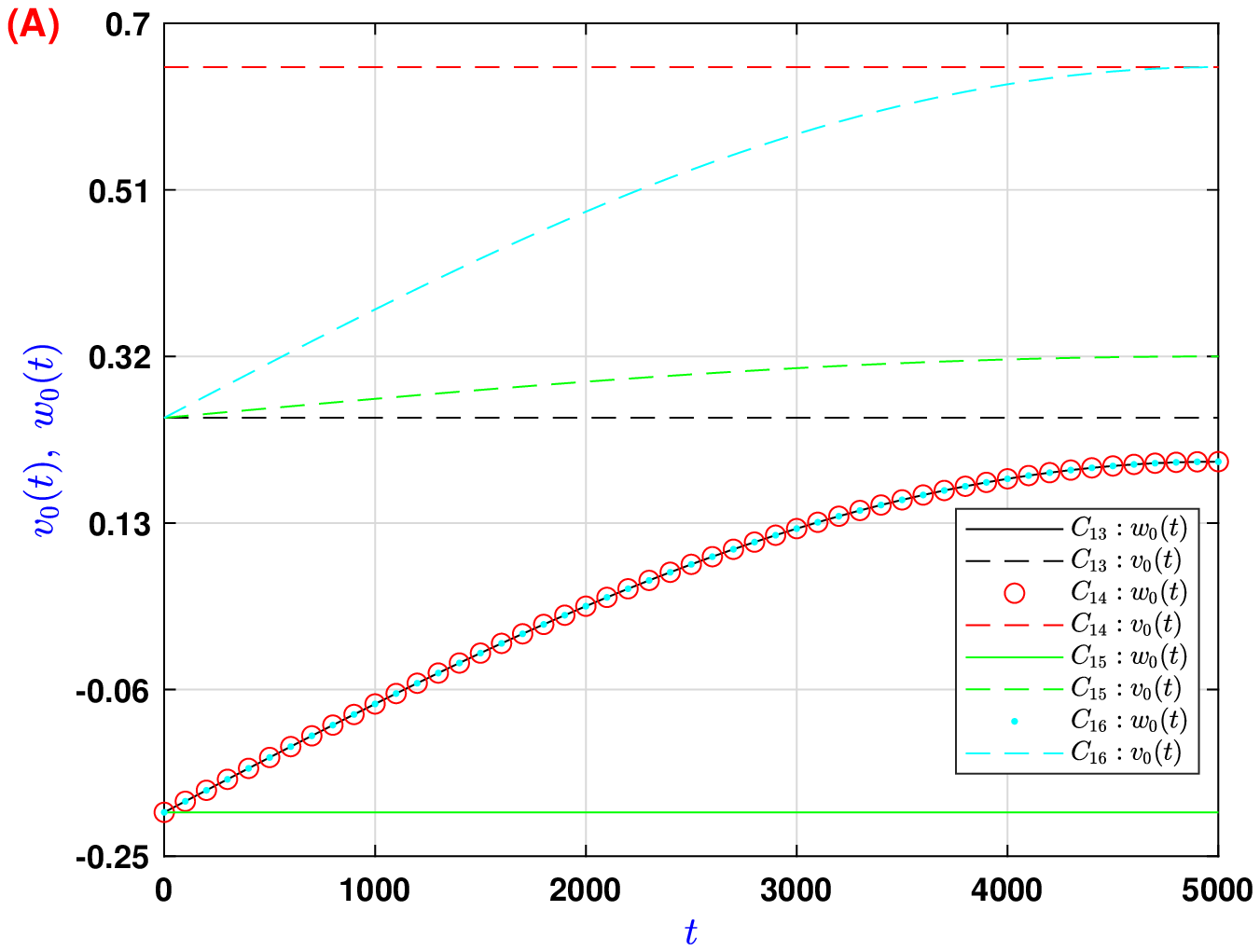}\includegraphics[width=5cm,height=3cm]{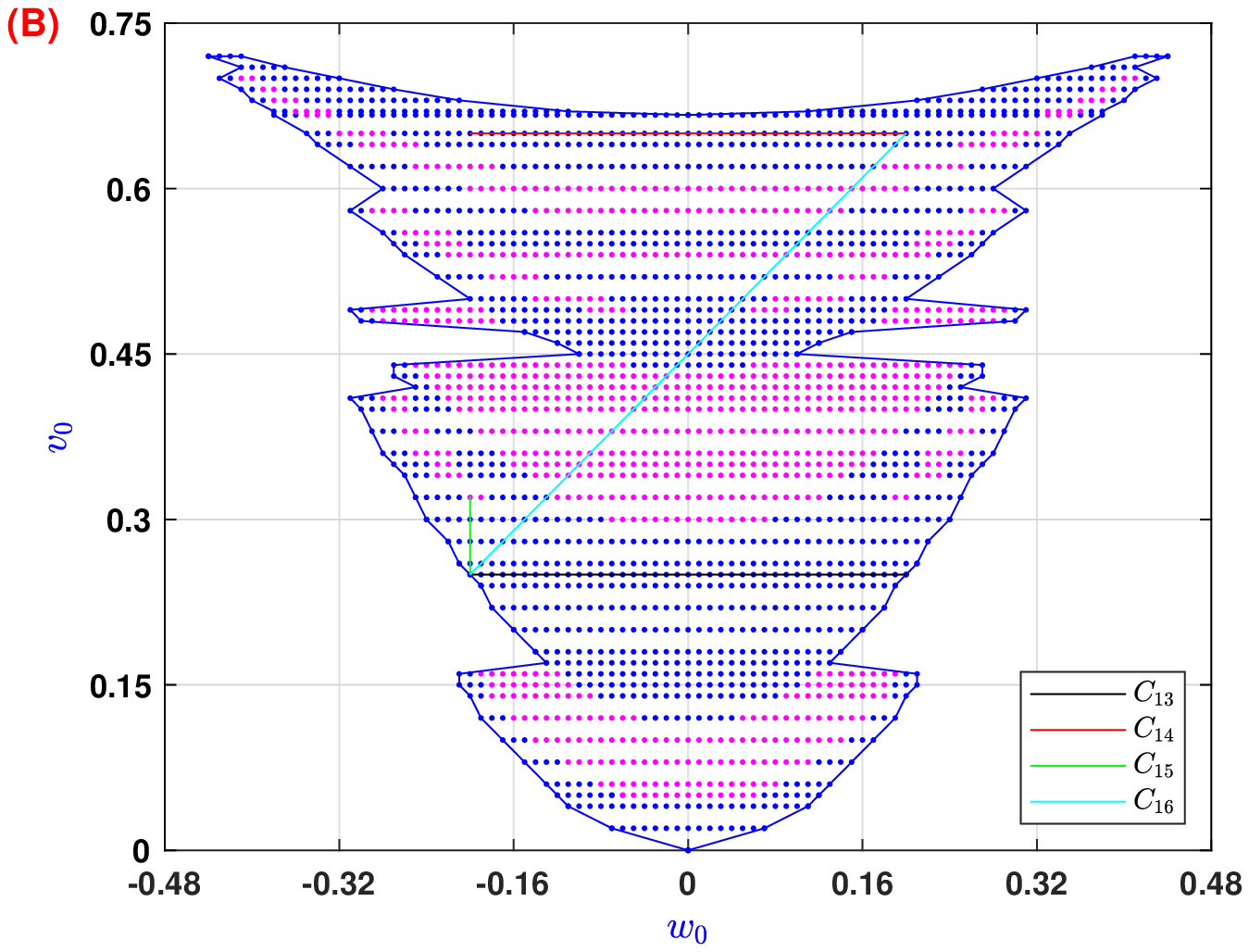}\\
	\includegraphics[width=5cm,height=3cm]{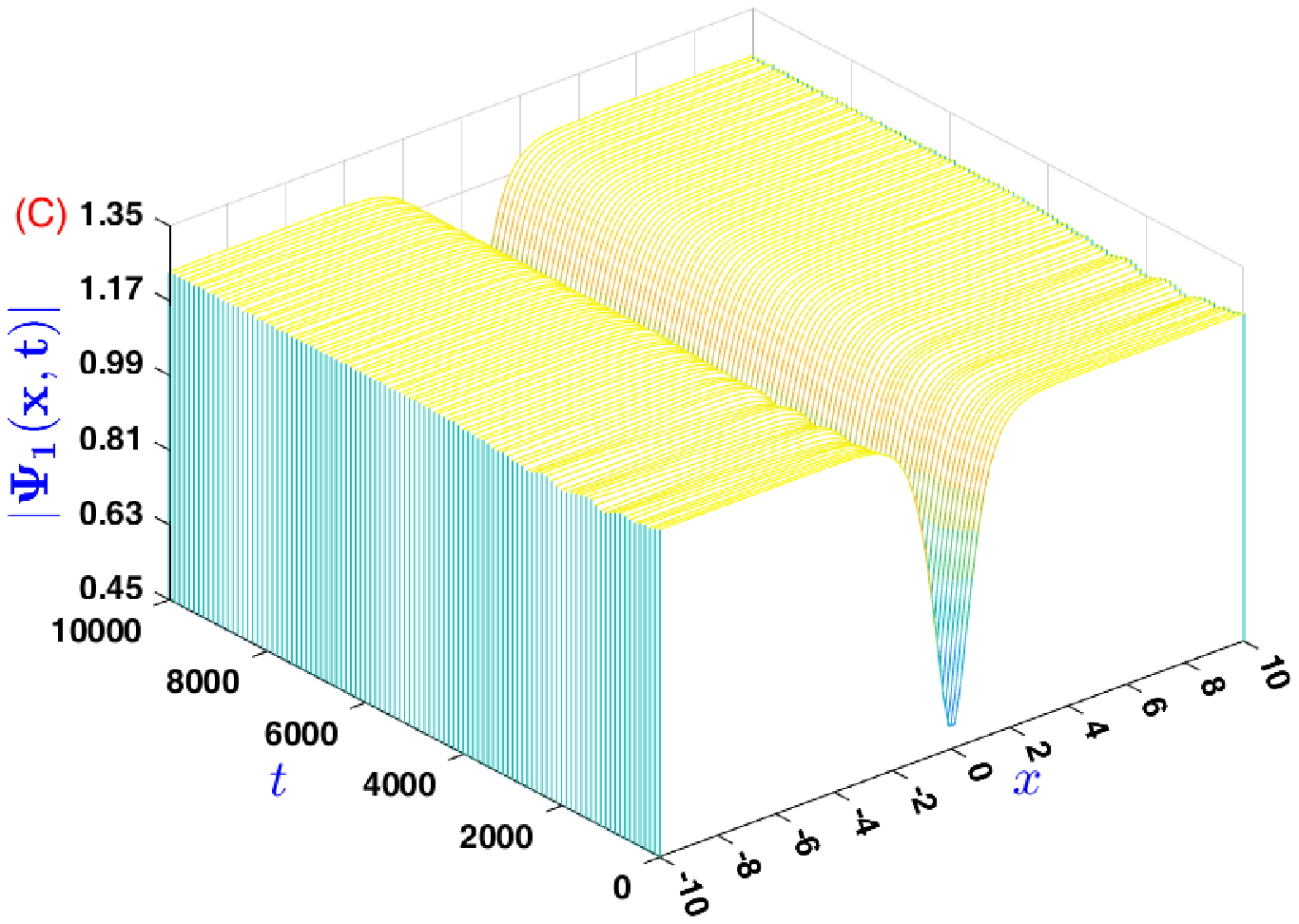}\includegraphics[width=5cm,height=3cm]{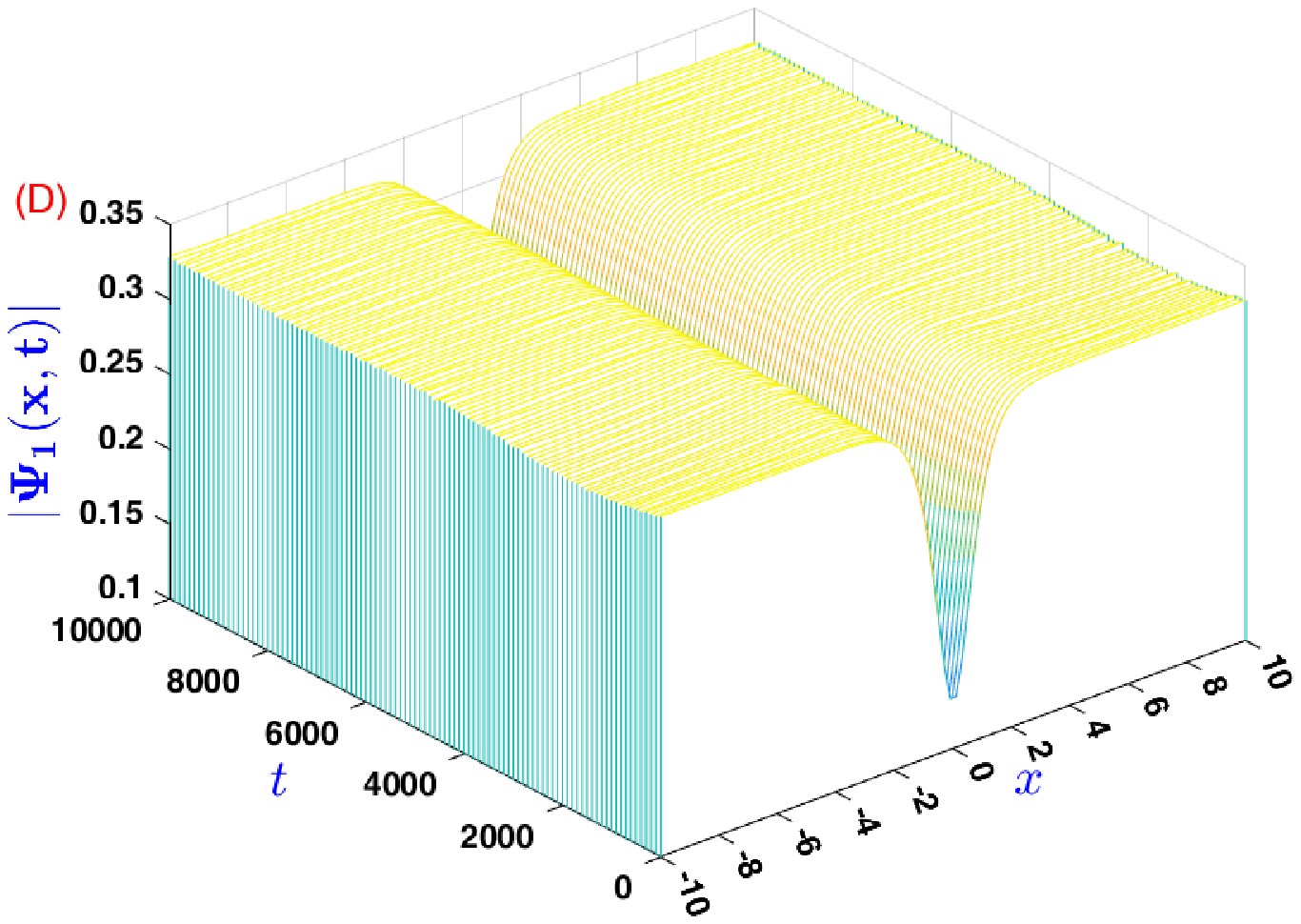}\\
	\includegraphics[width=5cm,height=3cm]{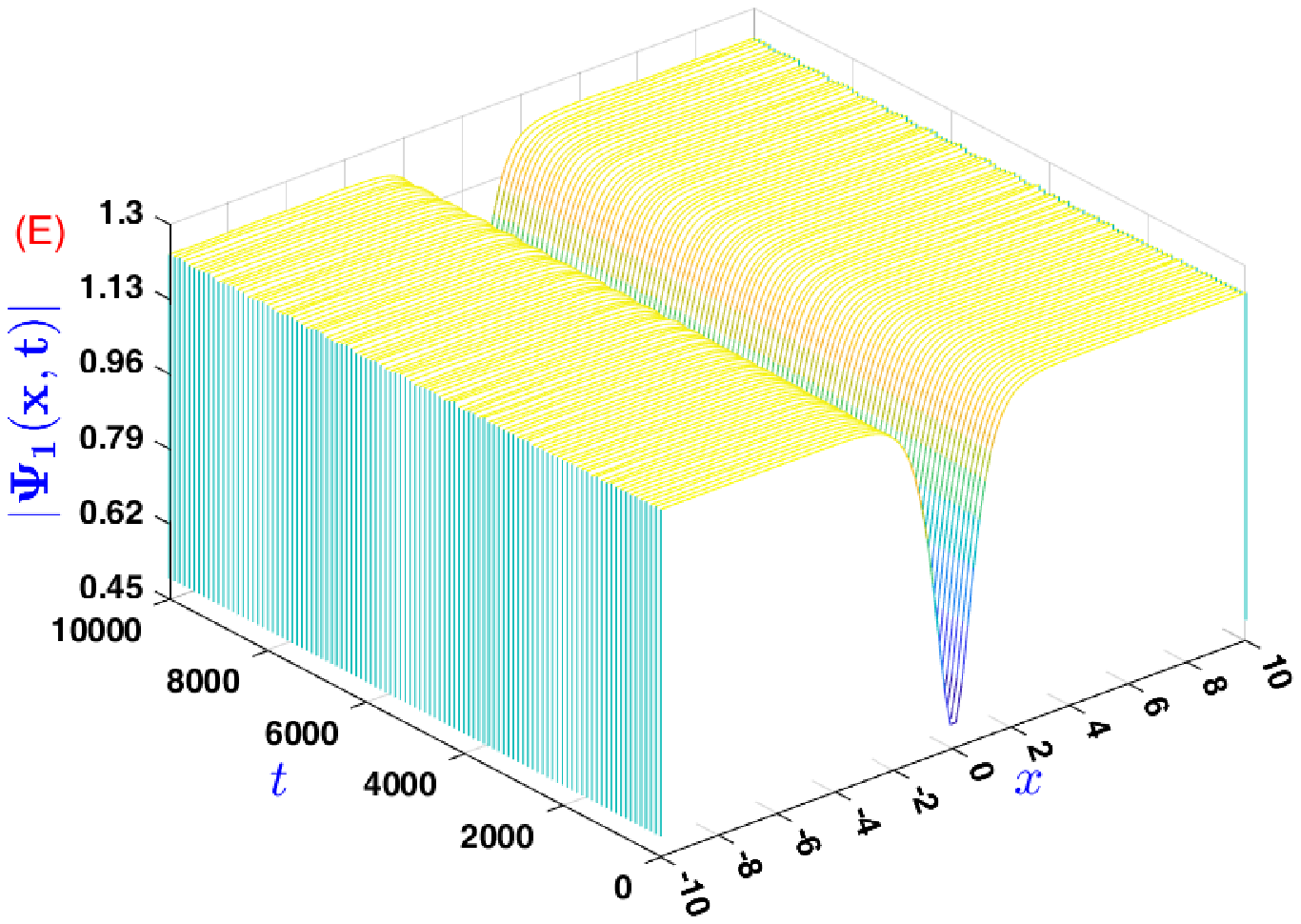}\includegraphics[width=5cm,height=3cm]{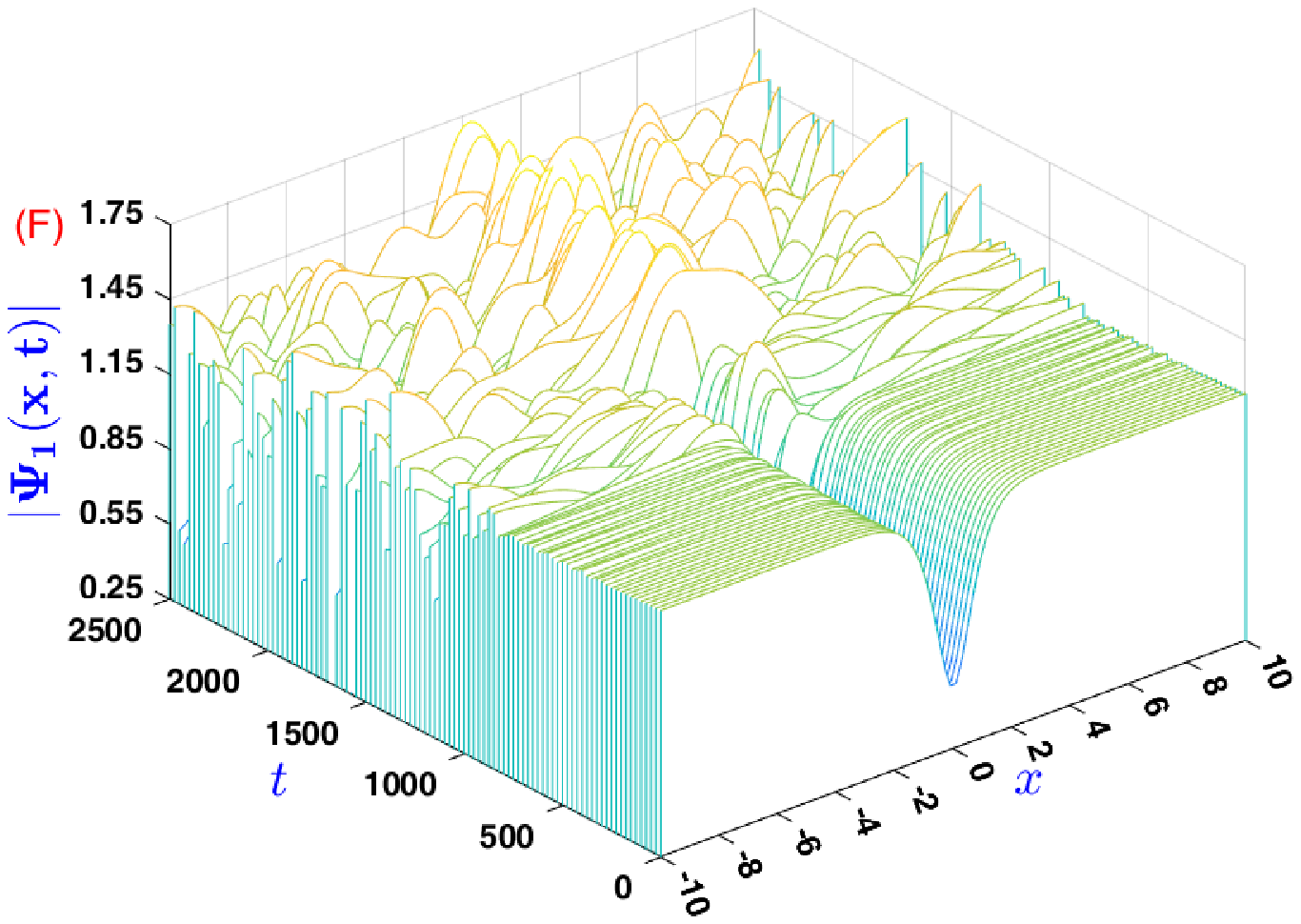}
	\caption{\label{Fig13.solution2g1excitationw0v0} (A) Shows time dependent potential amplitudes $w_0, v_0$ for different curves. (B) Shows that stable (\emph{blue dots}) and unstable (\emph{magenta dots}) modes of DS and four curves $C_i,~i=13-16$. Excitations of stable-unstable nonlinear delocalized state of (\ref{nlstExcitation}) with initial condition (\ref{solpsi1}), for $g=1$, $\b=1,\d=1$, (C) $v_{01}=0.25,v_{02}=0.25,w_{01}=-0.20,w_{02}=0.20$; (D) $v_{01}=0.65,v_{02}=0.65,w_{01}=-0.20,w_{02}=0.20$; (E) $v_{01}=0.25,v_{02}=0.32,w_{01}=-0.20,w_{02}=-0.20$; (F) $v_{01}=0.25,v_{02}=0.65,w_{01}=-0.20,w_{02}=0.20$.}
\end{figure}

In this case, time dependent potential amplitudes $w_0(t)$ and $v_0(t)$ are shown in Fig. \ref{Fig13.solution2g1excitationw0v0} (A), for four curves $C_{13}:P_1(-0.2,0.25)\longrightarrow P_2(0.2,0.25);~C_{14}:P_1(-0.2,0.65)\longrightarrow P_2(0.2,0.65);~C_{15}:P_1(-0.2,0.25) \uparrow P_2(-0.2,0.32);~C_{16}:P_1(-0.2,0.25)\nearrow P_2(0.2,0.65)$, and curves are shown in Fig. \ref{Fig13.solution2g1excitationw0v0} (B) and 
excitations of DS of (\ref{nlstExcitation}) with intial condition (\ref{solpsi1}) along these curves are shown in Fig. \ref{Fig13.solution2g1excitationw0v0} (C) - (F), for $g=\b=\d=1$ in $(w_0,v_0)$ plane. One can see that, curves $C_{13}$, $C_{14}$ and $C_{15}$ are passed through stable modes (\emph{blue dots}) but $C_{16}$ passes through some stable (\emph{blue dots}) and unstable modes (\emph{magenta dots}) and DS is stable along $C_{13}$, $C_{14}$, $C_{15}$ and unstable along $C_{16}$. Therefore, DS is stable in a disconneted $\mathcal{PT}$ unbroken region in $(w_0,v_0)$ plane.

\begin{figure}[h]
	\centering
	\includegraphics[width=5cm,height=3cm]{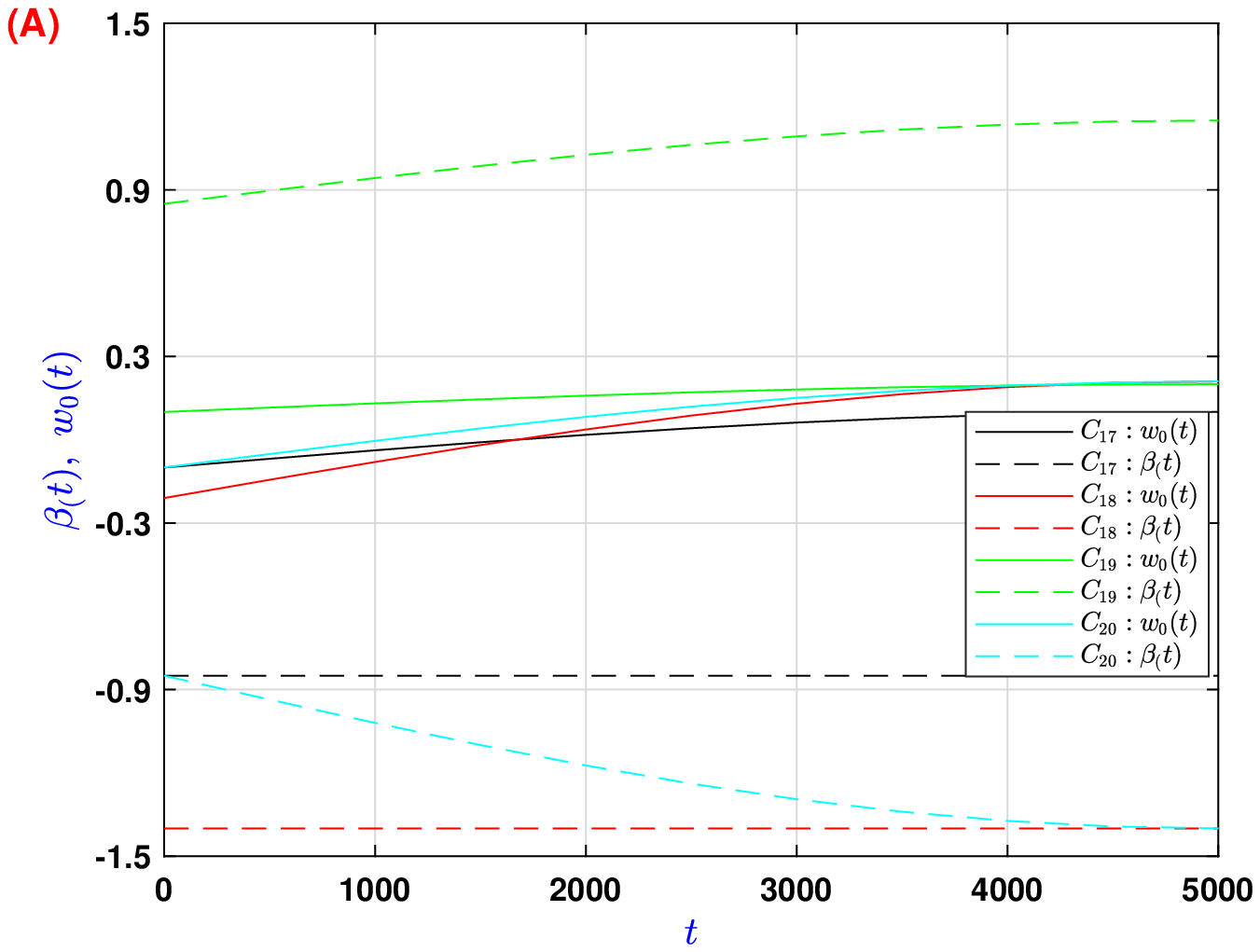}\includegraphics[width=5cm,height=3cm]{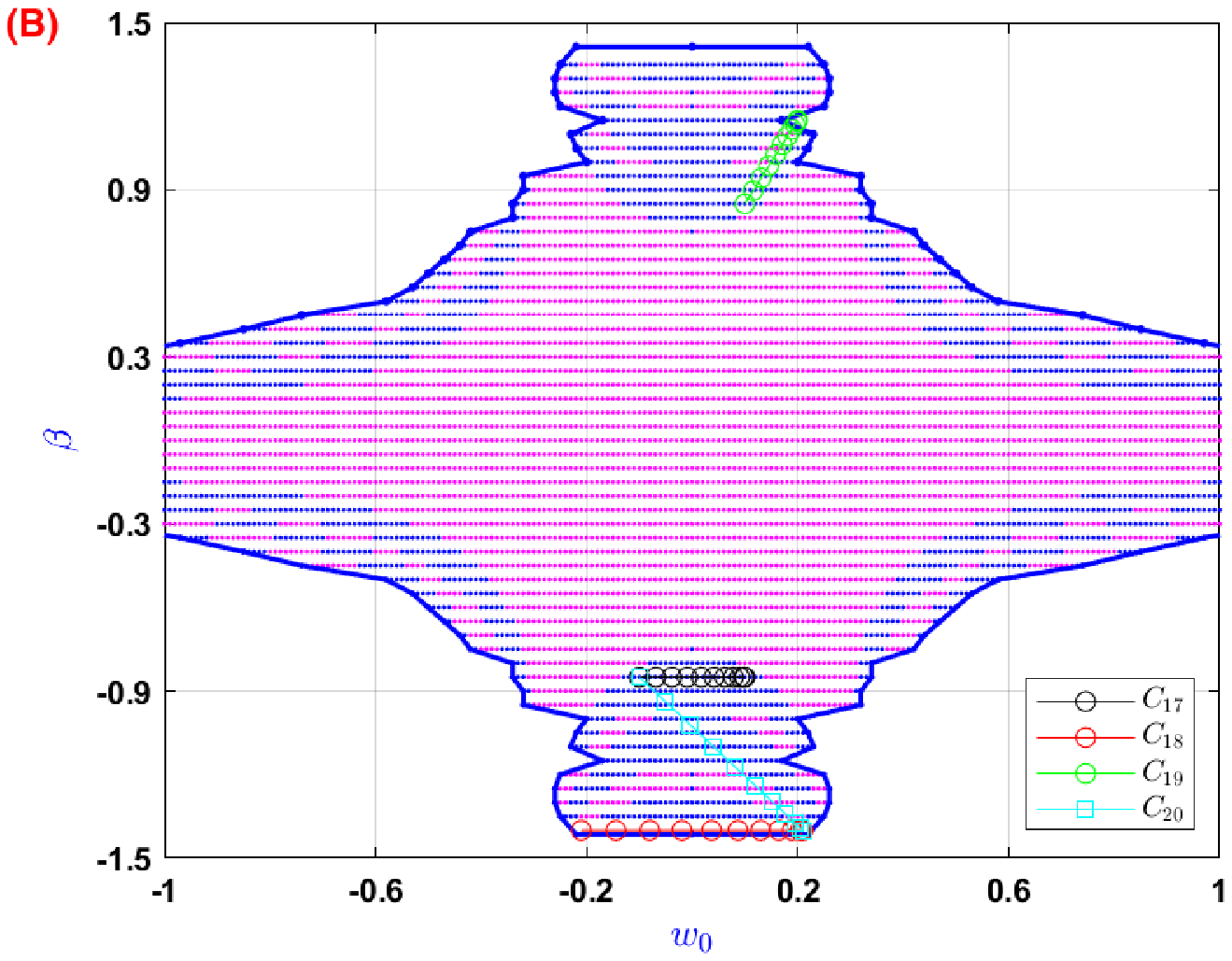}\\
	\includegraphics[width=5cm,height=3cm]{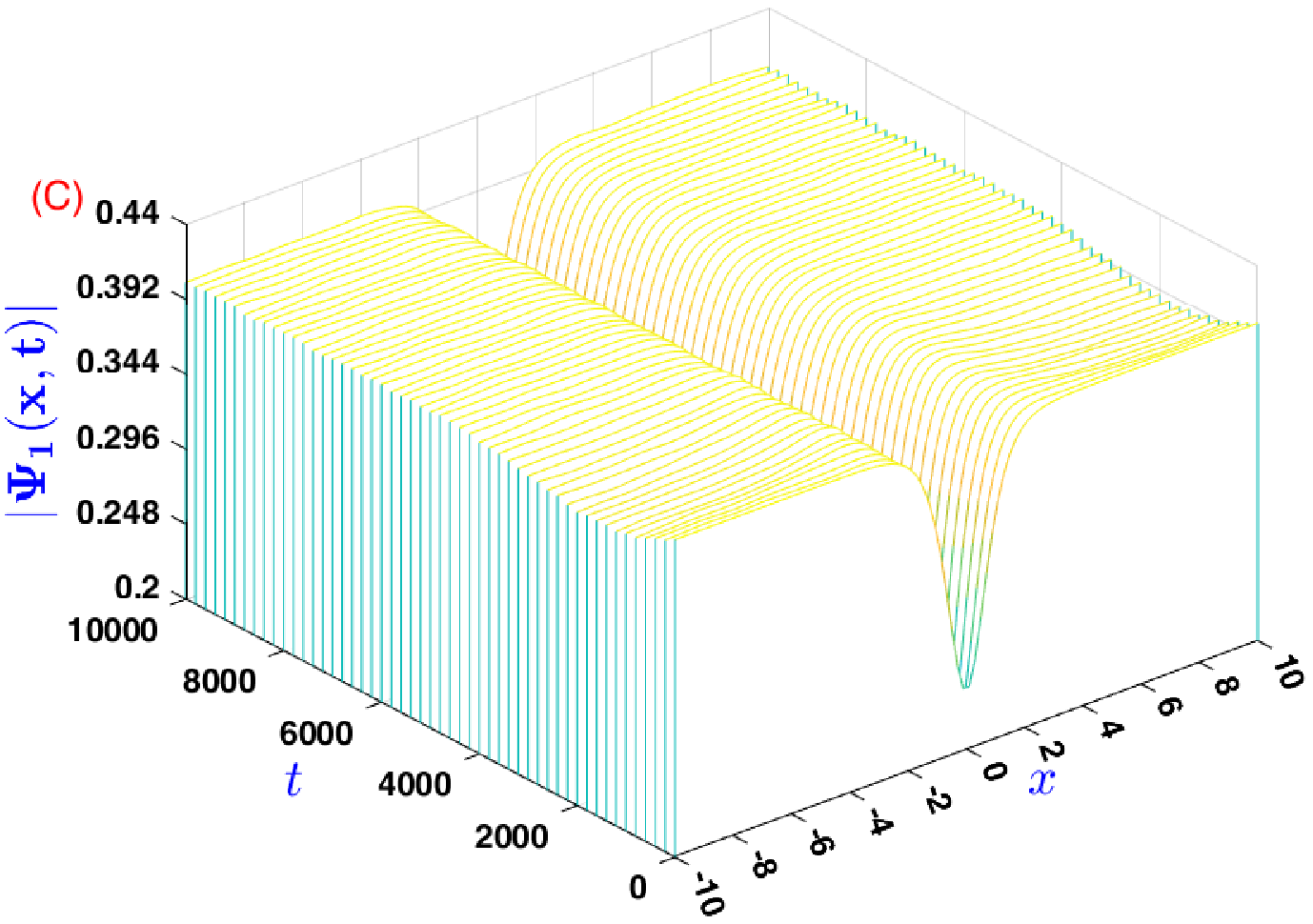}\includegraphics[width=5cm,height=3cm]{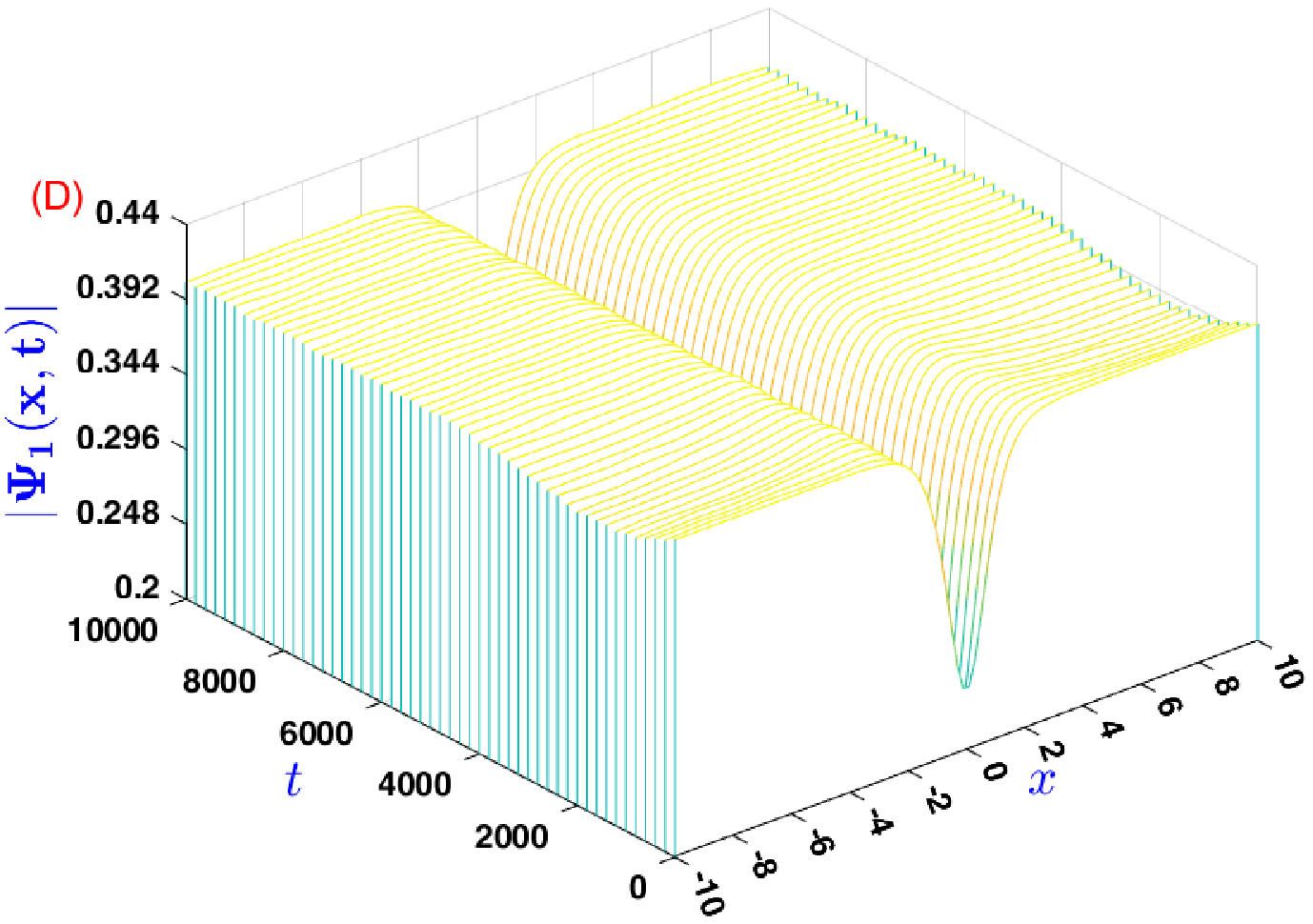}\\
	\includegraphics[width=5cm,height=3cm]{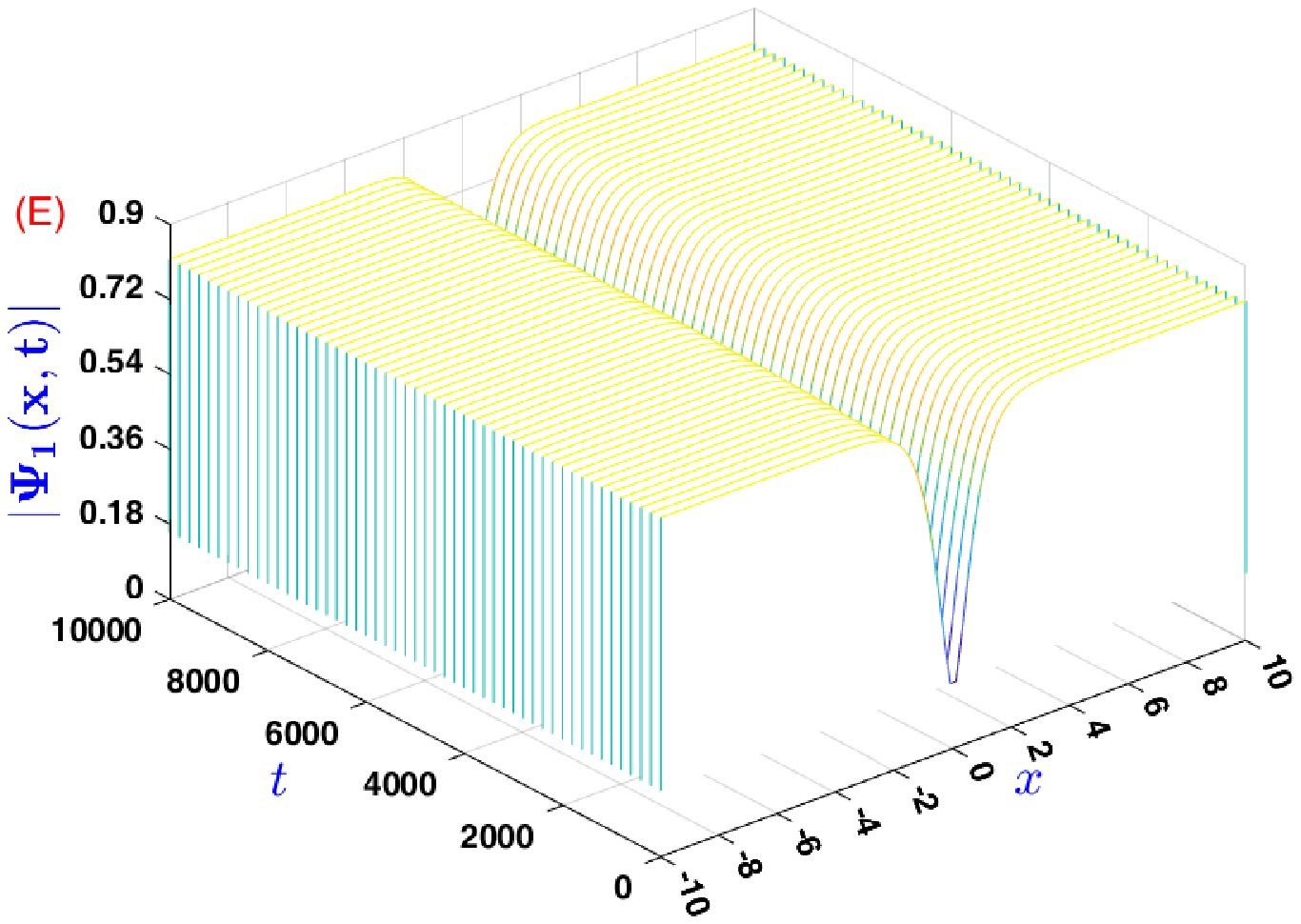}\includegraphics[width=5cm,height=3cm]{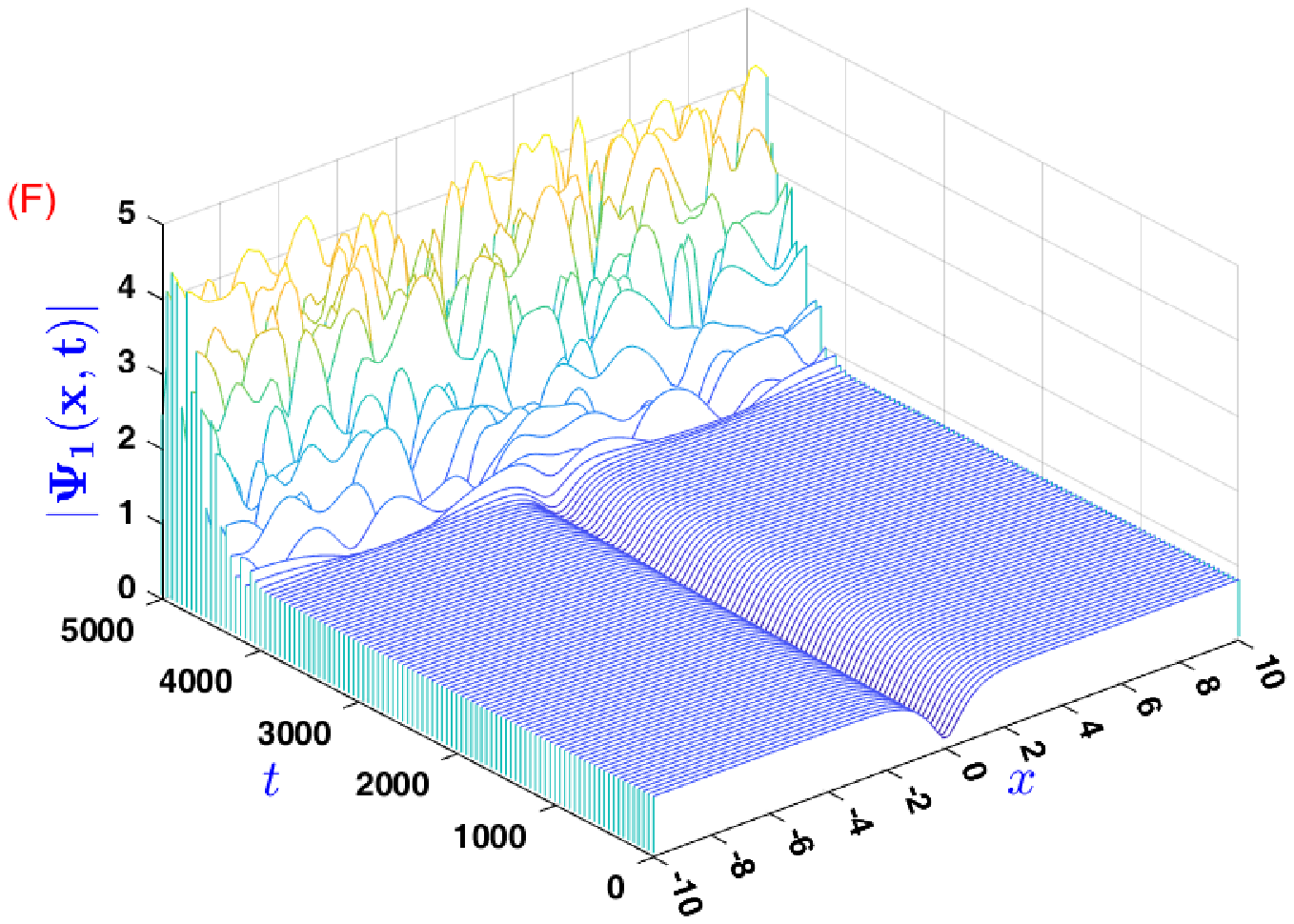}
	\caption{\label{Fig14.solution2g1excitationw0beta} (A) Shows time dependent potential amplitudes $w_0, v_0$ for different curves. (B) Shows that stable (\emph{blue dots}) and unstable (\emph{magenta dots}) modes of DS and four curves $C_i,~i=17-20$. Excitations of stable-unstable nonlinear delocalized state of (\ref{nlstExcitation}) with initial condition (\ref{solpsi1}), for $g=1$, $v_0=0.5,\d=1$, (C) $\b_{01}=-0.85,\b_{02}=-0.85,w_{01}=-0.10,w_{02}=0.10$; (D) $\b_{01}=-1.40,\b_{02}=-1.40,w_{01}=-0.21,w_{02}=0.21$; (E) $\b_{01}=0.85,\b_{02}=1.15,w_{01}=0.10,w_{02}=0.20$; (F) $\b_{01}=-0.85,\b_{02}=-1.40,w_{01}=-0.10,w_{02}=0.21$.}
\end{figure}

Similarly, excitations of DS of (\ref{nlstExcitation}) in $(w_0,\b)$ plane are evaluated along $(C_{17}:P_1(-0.1,-0.85)\longrightarrow P_2(0.1,-0.85);~C_{18}:P_1(-0.21,-1.4)\longrightarrow P_2(0.21,-1.4);~C_{19}:P_1(0.1,0.85)\nearrow P_2(0.2,1.15);~C_{20}:P_1(-0.1,-0.85)\searrow P_2(0.21,-1.4))$ are shown in Fig. \ref{Fig14.solution2g1excitationw0beta} (C) - (F), for $g=1, v_0=0.5,\d=1$ and curves are shown in Fig. \ref{Fig14.solution2g1excitationw0beta} (B) and the corresponding $w_0(t), \b(t)$ are shown in Fig. \ref{Fig14.solution2g1excitationw0beta} (A). One can see that DS is stable along $C_{17}$, $C_{18}$, $C_{19}$ and unstable for $C_{20}$, where the curve $C_{20}$ passes through stable and unstable modes of DS. Theofore, DS is stable in a disconnected $\mathcal{PT}$ unbroken region in $(w_0,\b)$ plane. 
\section{Conclusion}\label{sec5.con}
In this paper, a cubic NLSE with $\mathcal{PT}$-symmetric potential has been considered. A bright soliton and a bright-dark soliton interaction are generated in terms of potential amplitudes $v_0$ ,$w_0$ and $\b$. Depending on these parameters, bright-dark soliton interaction reduces to a BS if $w_0^2>\f{1}{4\b^2}$ and a DS, if $w_0^2<\f{1}{4\b^2}$. A phase transition occurs between BS and DS. In addition, $\mathcal{PT}$ broken-unbroken phase transition is shown with respect to $w_0, v_0, \b$ and also it is shown in $(w_0,v_0)$ and $(w_0,\b)$ planes. The linear stability analysis of BS and DS are investigated and obtained stable regions of BS in $\mathcal{PT}$ broken unbroken region and of DS in $\mathcal{PT}$ unbroken region.  
We have found that, (i) BS is stable in two connected regions $I_3\cup J_1\cup J_2$ (see Fig. \ref{Fig6.solution1g-1phasetransition} (A)) in $(w_0,v_0)$ plane and $I_7$ (see Fig. \ref{Fig6.solution1g-1phasetransition} (B)) in $(w_0,\b)$ plane for $g=-1$. 
(ii) BS is stable in two disconnected regions $J_1 \cup J_2$ (see Fig. \ref{Fig7.solution1g1phasetransition} (B))  in $(w_0,v_0)$ plane and $\cup_{i=3}^6 J_i$ (see Fig. \ref{Fig7.solution1g1phasetransition} (D), (E))  in $(w_0,\b)$ plane for $g=1$. 
(iii) DS is stable in a small connected region in $(w_0,v_0)$ plane such that $\frac{2+w_0^2}{3}\leq v_0\leq\frac{2+w_0^2}{3}+\epsilon$, where $0\leq\epsilon<0.1$, ${w_0}^2<\frac{1}{4}$, for $g=-1,\b=\d=1$. 
(iv) DS is stable in two disconnected regions $I_3$ (see Fig. \ref{Fig8.solution2g1phasetransition} (A)) in $(w_0,v_0)$ plane and  in $I_8$ (see Fig. \ref{Fig8.solution2g1phasetransition} (B)) $(w_0,v_0)$ plane for $g=1$.
(v) Solution (\ref{solpsi1}) is always unstable for $w_0^2> \f{1}{4\b^2}$.
All these regions are verified by eigenvalue problem (\ref{lameig}), then verified by direct numerical simulation, after that they are verified by exacitaions of BS and DS of time dependent nonlinear Schr\"odinger equation with time dependent potential amplitutes. The curves are taken with different length in different directions, which are defined in their respective domain. For examples only twenty excitaions are shown along twenty curves $C_i,~i=1-20$ and just twelve curves $C_i,~i=1-4,13-20$ are shown among them.
\vspace{-.5cm}
\section*{Acknowledgement}
Debraj Nath dedicates this article to the memory of his kind brother, late Raj Kumar Nath. Amiya Das gratefully acknowledges financial support from SERB-DST, Govt. of India (EEQ/2017/000150) and DST PURSE-II University of Kalyani.

\end{document}